\documentclass[runningheads]{llncs}

\usepackage[nolist]{acronym}

\usepackage{fancyhdr}
\usepackage{fancyvrb}

\usepackage{mathtools}

\usepackage{array}
\usepackage{float}
\usepackage{tikz}
\usepackage{supertabular}

\usepackage{multicol}
\usetikzlibrary{arrows,positioning}

\usepackage{breqn}

\usepackage{tabularx,ragged2e}
\newcolumntype{Y}{>{\centering\arraybackslash}X}

\usepackage{hyperref}
\usepackage{xcolor}
\usepackage{comment}
\usepackage{graphicx}

\usepackage{amssymb}
\usepackage{amsmath}
\usepackage{amsfonts}
\usepackage{enumitem}
\usepackage[normalem]{ulem}

\usepackage[linesnumbered,ruled,vlined]{algorithm2e}
\makeatletter
\newcommand{\removelatexerror}{\let\@latex@error\@gobble}
\makeatother

\usepackage{pgf-umlsd}
\usepgflibrary{arrows} 

\usepackage{listings}

\newcommand{\abs}[1]{\left\lvert{#1}\right\rvert}

\newif\ifcomments
\commentstrue

\let\oldnl\nl
\newcommand{\nonl}{\renewcommand{\nl}{\let\nl\oldnl}}

\newcommand{\todo}[1]{\textcolor{red}{[#1]}} 

\newcommand{\chris}[1]{\textcolor{blue}{CMS: #1}}

\newcommand{\adv}{\mathcal{A}}
\newcommand{\cdv}{\mathcal{C}}
\newcommand{\prob}[1]{\mathsf{Pr}\Big[ #1 \Big]}

\newcommand{\Exp}{\mathrm{Exp}}

\SetKwInput{ExecCon}{Execution Context}
\SetKwInput{Constants}{Constants}

\newcommand{\A}{\mathcal{A}}
\newcommand{\B}{\mathcal{B}}

\newcommand{\D}{\mathcal{D}}
\newcommand{\E}{\mathcal{E}}
\newcommand{\calL}{\mathcal{L}}
\newcommand{\K}{\mathcal{K}}
\newcommand{\calO}{\mathcal{O}}
\newcommand{\R}{\mathcal{R}}
\newcommand{\calS}{\mathcal{S}}

\newcommand{\KGen}{\mathsf{KGen}}
\newcommand{\WrapKey}{\mathsf{WrapKey}}
\newcommand{\UnwrapKey}{\mathsf{UnwrapKey}}
\newcommand{\ProvisionKey}{\mathsf{ProvisionKey}}
\newcommand{\EvictKey}{\mathsf{EvictKey}}
\newcommand{\VerifyToken}{\mathsf{VerifyToken}}
\newcommand{\EnableFBE}{\mathsf{EnableFBE}}
\newcommand{\Encrypt}{\mathsf{Encrypt}}
\newcommand{\Decrypt}{\mathsf{Decrypt}}
\newcommand{\efnode}{e{\normalfont \textit{f}}node}
\newcommand{\fnode}{{\textit{f}}node}

\newcommand{\enode}{enode}
\newcommand{\sigmaciph}{\sigma\mbox{-}ciph}
\newcommand{\sigmakey}{\sigma\mbox{-}key}
\newcommand{\fileciph}{\textit{f}s\mbox{\_}ciph}

\newcommand{\set}[1]{\{#1\}}
\def\getsr{\stackrel{\$}{\leftarrow}}
\newcommand{\get}{\leftarrow}

\newcommand{\ov}[1]{\overline{#1}}

\newcommand{\schemeacr}{MobileFBE}

\newcommand{\schemesymb}{\mathsf{MFBE}}
\newcommand{\mfbe}{\mathsf{MFBE}}

\begin{document}
\title{SoK: Untangling File-based Encryption on Mobile Devices}

\author{David Galindo\inst{1,2} \and Jia Liu\inst{2} \and
Chris McMahon Stone\inst{1} \and Mihai Ordean\inst{1}}

\institute{University of Birmingham, UK\\
\and
Fetch.ai\\
\email{\{d.galindo,c.mcmahon-stone,m.ordean\}@cs.bham.ac.uk, jia.liu@fetch.ai}
}

\maketitle

\begin{abstract}
File-based encryption (FBE) schemes have been developed by software vendors to address security concerns related to data storage. While methods of encrypting data-at-rest may seem relatively straightforward, the main proponents of these technologies in mobile devices have nonetheless created seemingly different FBE solutions. As most of the underlying design decisions are described either at a high-level in whitepapers, or are accessible at a low-level by examining the corresponding source code (Android) or through reverse-engineering (iOS), comparisons between schemes and discussions on their relative strengths are scarce. In this paper, we propose a formal framework for the study of file-based encryption systems, focusing on two prominent implementations: the FBE scheme used in Android and Linux operating systems, as well as the FBE scheme used in iOS. Our proposed formal model and our detailed description of the existing algorithms are based on documentation of diverse nature, such as whitepapers, technical reports, presentations and blog posts, among others. Using our framework we validate the security of the existing key derivation chains, as well as the security of the overall designs, under widely-known security assumptions for symmetric ciphers, such as IND-CPA or INT-CTXT security, in the random-oracle model.
\end{abstract}

%
%


\section{Introduction}

In the last decade mobile devices have become an integral part of modern life. As a result more and more sensitive data is accumulated on these devices which have become targets for third parties.

As a response to these potential dangers, the  smartphone industry has designed mechanisms to secure their customers devices, starting with securing access control with mandatory user authentication mechanisms, from biometrics and passcodes, all the way through to protecting the data stored on the devices using  file encryption systems. This has however led to a number of controversies such as the 2016 San Bernadino "FBI vs Apple" case in which the FBI demanded (unsuccessfully) from Apple to undermine their existing iOS security features by implementing backdoors that would allow government officials to access encrypted data under specific circumstances \cite{AppleSanBer,FBISanBer}.

Technical documents resulting from this case, as well as some more recent studies \cite{massdevdec}, have revealed significant gaps with respect to understanding the technical details of the security features implemented in these devices. This arises from the fact that often these features are only presented in whitepapers which brush over relevant technical details in favour of high level descriptions, e.g. \cite{ioswhitepaper,samsungtee}. Although alternative methods such as source code reviewing are in principle possible,  these are in most cases hindered by the manufacturers' preference towards closed source software. Thus, more often than not, true insight only becomes possible after significant reverse engineering campaigns \cite{samsungreveng}. In order to bridge these gaps multiple comprehensive studies about mobile device security have been done \cite{teufl2013ios,teufl2014android,smartphonesec}, however their main focus has been on providing details about the inner workings of the analysed schemes.

\subsection{Contributions}

In this paper we take a different approach, where we propose a new generic formalisation for File-Based Encryption (FBE) schemes with the goal of making existing and future schemes easier to analyse. In order to develop our formalisation we aggregate available information about FBE from multiple knowledge domains including whitepapers, technical documents, published source code and reverse engineering results (\cite{androidfbe,ioswhitepaper,androidslides,samsungtee,androidmetadata}, \cite{EXT4-whitepaper}, \cite{fscrypt_ker,keyutil,ext4encsource}, \cite{krsticios,meijer2018self,cooijmans2014analysis,androidmetadata}).

We carry out a formal analysis of the designs and implementations of FBE schemes, whilst also drawing up a comparison of the varying security guarantees that each implementation provides. More concretely, we provide:

\begin{enumerate}
    \item A formalisation of FBE schemes
    \item Generic adaptations using our formalism of the main FBE schemes used by Android and iOS, as reconstructed from multiple knowledge domains
    \item An up-to-date security analysis for the schemes used in Android and iOS
\end{enumerate}

In our analysis we target the main processes that enable FBE schemes to work, which we group in two categories: \textit{C1}) the key derivation methods, and root of trust placement and \textit{C2}) file encryption and decryption algorithms. For algorithms and methods in C1) we provide complete generic descriptions, in Sections \ref{sec:pre}-\ref{sec:keymanagement}, and specific analysis for both Android and iOS. We found that algorithms from this category have atypical implementations which mostly arise from hardware limitations or user interaction requirements. Therefore, where appropriate, we also consider functionality that depends on user inputs, such as passcodes or biometrics, to secure the keys that protect the memory storage of the device. Algorithms in \textit{C2}) follow the standard practices for FBE. We cover these in Section \ref{sec:ext4enc} for Android and *NIX systems only, due to limited availability of appropriate low-level documentation for iOS.
We argue that our findings will allow manufacturers to accurately pinpoint the components which have the biggest impact on the security of the device and allocate more resources towards securing these. Conversely, this work also allowed us to identify attack scenarios which do not directly translate to compromising the whole security of the device.

\section{File-based encryption security}

In this section we will give the syntax for a generic $\schemeacr$ scheme, followed by a definition of the corresponding security policies.

\subsection{Scheme definitions.}

\begin{definition}
A \emph{\schemeacr{} scheme\footnote{We sometimes use the terms \textit{scheme} and \textit{protocol} interchangeably.}} with a defined set of policies $\mathcal{P}$ is a tuple of eight polynomial-time algorithms  $\schemesymb{}=\allowbreak(
    \allowbreak\KGen,
    \allowbreak\WrapKey,
    \allowbreak\UnwrapKey,
    \allowbreak\ProvisionKey,
    \allowbreak\EvictKey,
    \allowbreak\VerifyToken,
    \allowbreak\Encrypt,
    \allowbreak\Decrypt
)$ such that:
	
	\begin{description}
		\item[$k \leftarrow \KGen(1^\lambda):$] is a probabilistic algorithm for generating $\schemesymb{}$ class keys. It takes as input the security parameter $\lambda$ and outputs a random value $k$ of length $\lambda$.
		
		\item[$\{\phi, \overline{pol}\} \leftarrow \WrapKey^\sigma(k,\omega,pol):$] is a probabilistic algorithm that takes as input an unencrypted key $k$, an encrypted wrapping key $\omega$, and a unbound policy $pol$. The policy $pol$ contains information related to the intended usage of key $k$ and the ciphers used to encrypt it. The algorithm returns $\phi$, an encryption of $k$ with respect to $\omega$, and $\overline{pol}$ the policy $pol$ bound with the wrapping key $\omega$.
		
		\item[($k,\bot) \leftarrow \UnwrapKey^\sigma(\phi,\overline{pol}):$] is a deterministic algorithm that takes as input an encrypted key $\phi$ and a bound policy $\overline{pol}$. It returns a file encryption key $k$ obtained by decrypting $\phi$ with the wrapping key bound in $\overline{pol}$. Depending on the cipher used in the policy $\bot$ may returned if the decryption is not successful.
		
		\item[$kid \leftarrow \ProvisionKey(k):$] is a deterministic algorithm to provision keys for their usage. It takes as input a key $k$ and outputs an identifier $kid$ of the  key thus installed.
		
		\item[$\{1,\bot\} \leftarrow \EvictKey(kid):$] is a deterministic algorithm used to discard keys. It takes as input the key identifier $kid$ to be discarded from memory and outputs $1$ if the operation was successful or $\bot$ otherwise. 
		
		\item[$\{1,\bot\}  \leftarrow \VerifyToken(authToken):$] is a deterministic algorithm used to verify the validity of a cryptographically generated token $authToken$. It returns $1$ if the token verification was successful or $\bot$ otherwise.

		\item[($e{\normalfont \textit{f}}node,\bot) \leftarrow \Encrypt({\normalfont \textit{f}node},kid):$] is a probabilistic algorithm to encrypt a FBE filesystem node (e.g. file or file directory). It takes as input the node, ${\normalfont \textit{f}}node$, and a key or key id, $kid$; outputs its encrypted counterpart, $efnode$. It may return $\bot$ if the identifier $kid$ is not valid.
		
		\item[($\fnode,\bot) \leftarrow \Decrypt(\efnode,kid):$] is a deterministic algorithm to decrypt a FBE encrypted node. It takes as input the encrypted node, $\efnode$, and a key or key id, $kid$; outputs its decryption, $\fnode$. It may return $\bot$ if the decryption was not successful (i.e. invalid key or $kid$).

	\end{description}
\end{definition}

\section{Preliminaries}
\label{sec:pre}
In this section we focus on the syntax of security policies and how we use them to accurately describe existing key policies used in mobile operating systems. 

\subsection{Generic syntax of security policies}
\label{sec:gen_policy}

We use the notion of a security \textit{policy} to express (1) the usage scenarios allowed for keys in a \schemeacr{} scheme, and (2) the security parameters required for the management of these keys. These policies describe how and when keys are built, provisioned and stored. The security parameters within the policy also dictate relationships between keys. This allows us to capture details such as whether a key is  cryptographically bound to an external process or device, e.g. the Secure Enclave Processor in iOS. As such, a generic \textit{policy} is defined as a tuple of four parameters:
\begin{multline*}
    pol = [cipher, wrappingKey, authToken, usage]
\end{multline*}

We distinguish two types of policies: unbound policies $pol$  which are associated to new, unprotected file encryption keys, i.e. $(k, pol)$ and bound policies $\overline{pol}$ associated to protected file encryption keys, i.e. $(\phi,\overline{pol})$. The unbound policies do not yet contain any cryptographic material specific to a device or credential, such as wrapping keys or authentication tokens. When platform specific cryptographic material is added to the policy, the policy becomes bound to said platform and is denoted as $\overline{pol}$. The keys and parameters of the bound policy $\overline{pol}$ are used to protect the file encryption key associated to the policy (i.e. $\phi$) and restrict use of that key to that specific platform.

\smallskip\noindent\emph{The $cipher$ parameter.} This parameter specifies the cipher and parameters that were used to encrypt the key $k$ to which the policy is associated. We use standard notation to describe this (e.g. AES-256-GCM: AES cipher with a key of length 256bits and a GCM block chaining mode).

\smallskip\noindent\emph{The $wrappingKey$ parameter.} It refers to the ciphertext that contains the private key under which the key $k$ associated to a policy is encrypted. On platforms that have access to a trusted execution environment (TEE) the wrapping keys are encrypted with hardware bound keys that are exclusively stored inside the TEE. This ensures that recovery of the file encryption key cannot be done without access to the TEE. This also enforces verification of the policy inside the TEE. For platforms lacking a TEE wrapping keys are still used as they are part of the code base, but they will not provide any additional security benefits. It is worth noting that mobile platforms without TEEs are rare in practice nowadays.

\smallskip\noindent\emph{The $authToken$ parameter.} The authToken or authentication token is a platform specific, optional cryptographic value that encodes the user contribution (e.g. password, PIN, fingerprint reader output) in the form of a specific cryptographic hash (in Android), or a password derived key (in iOS). If the value is null then decrypting the key file encryption key $k$ can be done transparently, without any input from the user.  We discuss this further in the next section.

\smallskip\noindent\emph{The $usage$ parameter.} This final parameter is used to indicate the purposes and restrictions associated with the use of a key.

\subsection{The authToken}
\label{sec:authtokens}

\noindent\textit{Android.} In Android, the user input, e.g. the user passphrase or PIN, is encoded as a data structure refered to as \textit{AuthToken} \cite{androidauth}. The \textit{AuthToken} token generation and verification consists of two parts and is managed by the Gatekeeper trustlet in the TEE and a counterpart service running on the insecure application processor. During the enrolment phase (e.g. on first boot) a 64-bit user secure identifier, named \textit{AuthToken HMAC Key}, is created by Gatekeeper trustlet and is then shared between all the relevant TEE components, namely the  Gatekeeper, Keymaster and biometric trustlet. On subsequent operations which require user authentication, the Gatekeeper service running on the application processor will generate AuthTokens which need to be included as part of any requests. The submitted tokens are verified inside the TEE by the Keymaster trustlet which handles key management and verifies tokens related to FBE keys. Rate limiting of authentication attempts, based on time-delays as well as token freshness and correctness are enforced by the TEE trustlet \cite{androidauth}.

\medskip

\noindent\textit{iOS.} A similar functionality for handling user input exists in iOS, however it is enforced through encryption rather than MAC verification. A wrapping key $k_{master\_key}$ is derived inside the secure enclave processor (SEP) from the user input using a key derivation function. The derivation process enforces time-delayed rate limiting and, as such, the $k_{master\_key}$ will have the same security protections against brute-force attacks as the Android AuthToken equivalent \cite{krsticios}.

Due to the similarity in functionality, security properties and usage,  we refer to both the Android's \textit{AuthToken} and iOS's $k_{master\_key}$ with the generic term \textit{authToken} throughout the rest of the paper, and we make the distinction where relevant.

\subsection{Android policies}

Since the introduction of FBE in Android 7.0, there has been support for two main encryption contexts, called Device Encrypted (DE) and Credential Encrypted (CE). The former of the two protects data up until the point at which a successful Verified Boot\footnote{Verified boot guarantees the integrity of the Android OS software with hardware backed root of trust binding \cite{androidverboot}.} has taken place. The latter, which is the default storage location, makes data available once a user has unlocked the device for the first time, up until the device has been shut down \cite{androidfbe}. Data in both the DE and CE storage locations is encrypted with keys that are protected with trusted hardware. In Android, the trusted hardware consists of a Key management applet, named Keymaster, which runs in a secure enclave in ARM TrustZone.

\smallskip\noindent\emph{The $DE_{s}$ policy.} The device encryption context is split into two policies.  The first policy is called $DE_{s}$, and describes when and how system data is encrypted. This policy protects storage areas that  are associated with the device and that are common for all enrolled users. This includes, for example, Wi-Fi credentials and Bluetooth pairing data \cite{androidslides}. Using our syntax, the policy $DE_{s}$ associated with the key $\phi_{DEs}$ is:
\begin{multline*}
    \footnotesize
    DE_{s} = [\text{\textit{AES-256-GCM}}, \omega_{DEs}, null, DeviceDataAfterBoot],
\end{multline*}
\noindent where $\omega_{DEs}$ is the hardware bound ciphertext of the key used to protect $\phi_{DEs}$. Keys associated with this policy do not require any user input in order to be made available, and as such the authentication token is set to $null$.

\smallskip\noindent\emph{The $DE_{u}$ policy.} The second DE policy is the per-user context policy $DE_{u}$. Data encrypted under this policy is user specific, but is made available along with $DE_{s}$ data after boot without requiring any contribution from the user to enable usage. It includes data such as alarms, wallpapers and active ring tones \cite{androidslides}. There is a separate $DE_{u}$ master key for each enrolled user, where $u$ is the user id. Cryptographically speaking, the $DE_{u}$ policy is identical to $DE_{s}$, the only difference being the usage parameter:
\begin{multline*}
    \footnotesize
    DE_{u} = [\text{\textit{AES-256-GCM}}, \omega_{DEu}, null, UserDataAfterBoot],
\end{multline*}

\noindent where $\omega_{DEu}$ is a unique hardware bound ciphertext specific to each $\phi_{DEu}$.

\smallskip\noindent\emph{The $CE_{u}$ policy.} The $CE_{u}$ policy describes the credential encrypted context in Android. This is also a per-user encryption context. However, unlike $DE_{u}$, this context ensures data is protected  until after first authentication. This is the default policy for all user data, and includes files like documents, photos and app data. Similar to the $DE_{u}$ policy, there is a separate $(\phi_{CEu}, CE_u)$ key-policy pair for each user. Because files encrypted in this context are only decrypted after a successful user authentication, a verified authentication token $AuthToken$ is required by the Keymaster element to unwrap and install the $CE_u$ key. The policy can  be represented as:
\begin{multline*}
    \footnotesize
    CE_{u} = [\text{\textit{AES-256-GCM}}, \omega_{DEu}, \\AuthToken, UserDataAfterAuth]
\end{multline*}

\noindent where $\omega_{CEu}$ is a unique hardware bound ciphertext specific to each $\phi_{CEu}$, and $AuthToken$ is a cryptographic hash generated inside the TEE.

\smallskip\noindent\emph{Other keys.} Support for metadata encryption has been introduced since  Android 9. Under the current implementation model, data that is not already protected by the $DE$ or $CE$ keys, (e.g. file sizes, permissions) is considered to be metadata and is encrypted using a separate key, which we denote $k_{metadata}$. The metadata protection is implemented as a supplementary full disk encryption layer, i.e. an adaptation of Android 5.0 full disk encryption \cite{androidmetadata} that has been previously analysed in \cite{teufl2014android}.

\subsection{iOS policies}

Apple introduced its implementation of hardware-backed file-based encryption with iOS 4 in 2010 under the name \textit{Data Protection}. In the latest version of Data Protection, four primary encryption contexts are supported --- classes A to D. In reverse order, ClassD is the Android $DE_{s}$ equivalent and ClassC is the Android $CE_{u}$ equivalent. Unlike Android, iOS does not support cryptographic separation of per-user file system content (i.e. per-user $DE_{u}$), nor does it support multiple user profiles per device (i.e. more than one $CE_{u}$) \cite{ioswhitepaper,iosmultiuser}. Two additional policies are supported in iOS: ClassB, which allows writing encrypted data but not reading whilst a device is locked; and ClassA, which evicts corresponding keys from memory when a device is locked after a successful authentication \cite{ioswhitepaper}. All keys in this scheme are bound to the Secure Enclave Processor (SEP). Like Android, all class keys are wrapped using AES-256-GCM \cite{belenko2014}. 
However, unlike Android, in iOS all keys are generated and stored inside the SEP and all relevant encryption and decryption operations are run either by the SEP or by the specific hardware which intermediates access to the data (e.g. the storage controller). According to Apple, keys are never released into the main memory of controlled by the application processor \cite{ioswhitepaper}. Additionally, in iOS class keys are used not only to encrypt files but also to encrypt other class keys which have a more restrictive usage. Below we show how our proposed syntax can be used to model the iOS security policies.

\smallskip\noindent\emph{$ClassD$ policy.} Data protected under this policy consists primarily of system files needed for boot, but also Voicemail, Bluetooth and iMessage keys, VPN certificates, etc. As this is one of the most permissive keys in iOS it is only protected with the hardware key of the TEE (i.e. a unique 256bit AES key, fused into the SEP during manufacturing). The ClassD policy associated to the encrypted key $\phi_{clsD}$ is:
\begin{multline*}
    \footnotesize
	clsD = [\text{\textit{AES-256-GCM}}, null, null, FileProtectionNone]
\end{multline*}

We represent the fact that the key is directly bound to the TEE and that no additional ciphertext is required to decrypt this key (aside from the hardware key of the SEP) by setting both the wrapping key value and the user contribution, $authToken$, to null.

\begin{figure*}[ht]
    \centering
    \begin{minipage}{0.46\textwidth}
        \includegraphics[height=120pt]{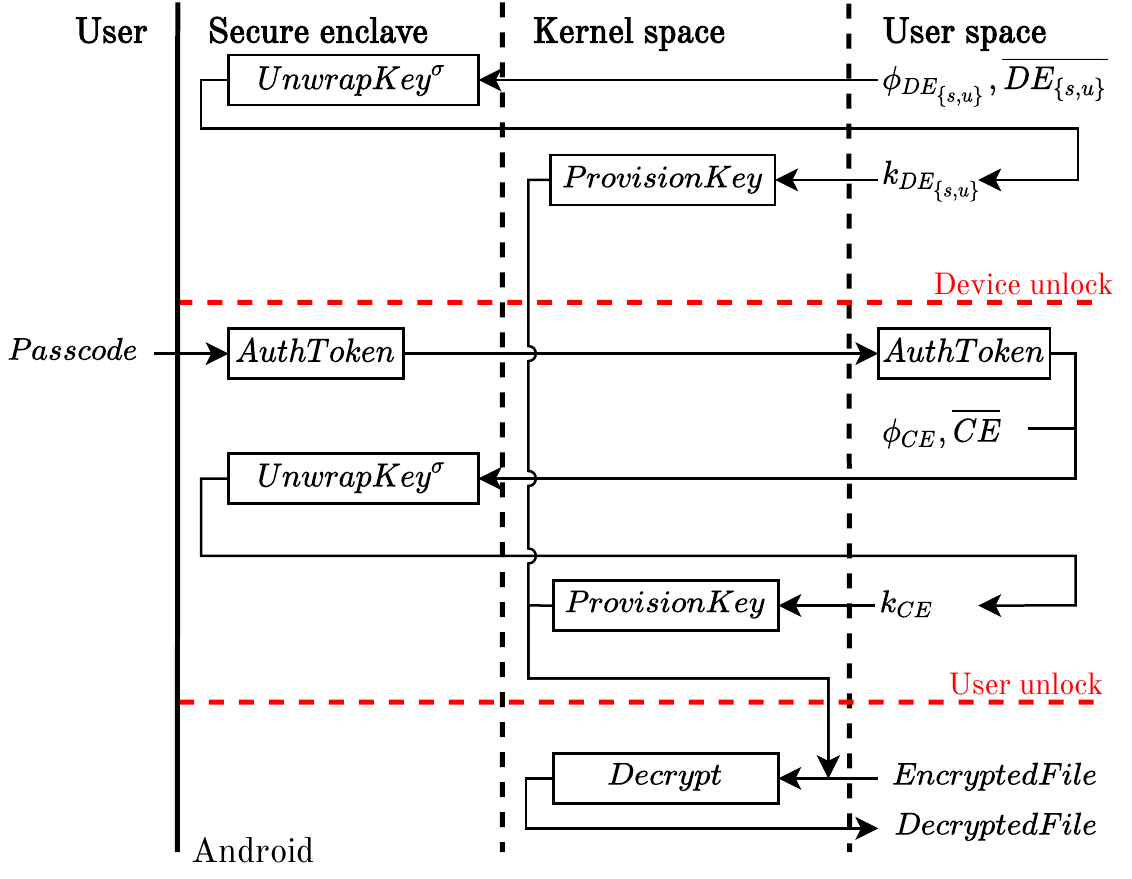}
    \end{minipage}
    \hspace{5pt}
    \begin{minipage}{0.46\textwidth}
        \includegraphics[height=120pt]{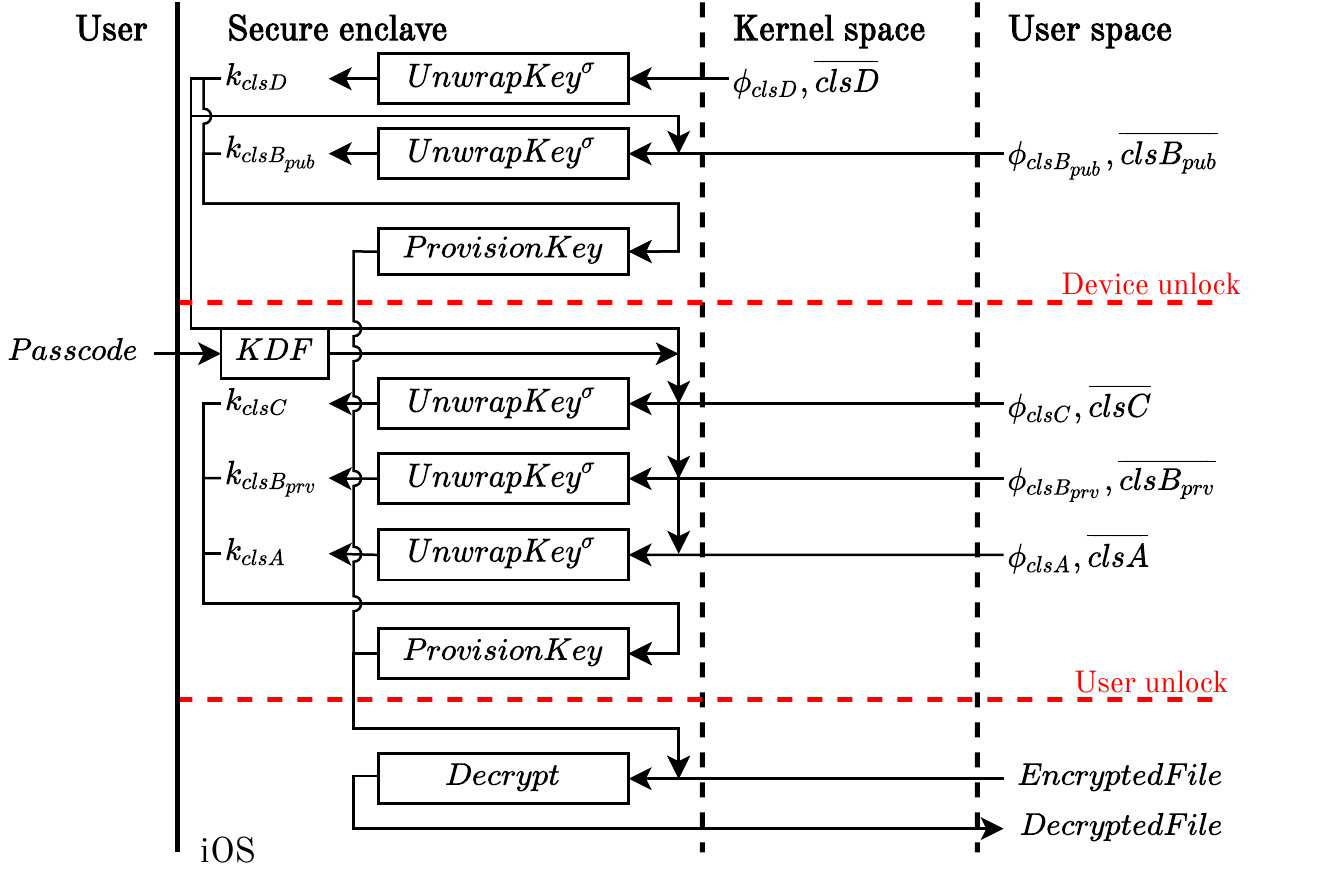}
    \end{minipage}
    \caption{Boot, Unlock and File Decryption stages on Android (left) and iOS (right) platforms are depicted here. Note that the process starts in the top right corner in each diagram. In Android a user space algorithm loads the keys from the system partition and starts the interaction with the secure enclave. In IOS the process is started when the kernel loads the wrapped $\phi_{clsD}$ key from a kernel partition. The other keys are subsequently retrieved from the user partition.}
    \label{fig:mfbe}
\end{figure*}

\smallskip\noindent\emph{$ClassC$ policy.} This is the default encryption policy for system and user apps and user data. Like the $CE_u$ class in Android, this class key is made available after first authentication, and not evicted when locking the device. The key is protected by $k_{clsD}$ and requires an authentication token consisting of a password derived key referred to as $k_{master\_key}$ \cite{ioswhitepaper}. Using our syntax we describe the policy associated to a ClassC key $\phi_{clsC}$ as:
\begin{multline*}
    \footnotesize
	clsC = [\text{\textit{AES-256-GCM}}, \phi_{clsD}, k_{master\_key},
	\\FileProtectionCompeteUtilFirstUserAuthentication]
\end{multline*}

\smallskip\noindent\emph{$ClassB$ policy.} This policy facilitates the ability to write encrypted data when the device is locked, whilst at the same time prevent reading/access of the same data using public key cryptography. This functionality is useful, for example, when an e-mail attachment is downloading in the background \cite{ioswhitepaper}. In order to use our syntax to describe this policy we have to split it into two parts, the public key  and private key parts. The public key of the policy is made available once the device has booted (immediately after the ClassD key $k_{clsD}$ as it is encrypted with it). The private key part, however, is only released after first authentication, together with  the ClassC key. Therefore the two policies for describing the ClassB policy are:
\begin{multline*}
    \footnotesize
	clsB_{pub} = [\text{\textit{AES-256-GCM}}, \phi_{clsD}, null, \\FileProtectionCompleteUnlessOpen]
\end{multline*}
\vspace{-25pt}
\begin{multline*}
    \footnotesize
	clsB_{priv} = [\text{\textit{AES-256-GCM}}, \phi_{clsD}, k_{master\_key}, \\FileProtectionCompleteUnlessOpen]
\end{multline*}

We again capture the user contribution requirement in the decryption of the ClassB private key part by setting the $authToken$ parameter.

\smallskip\noindent\emph{$ClassA$ policy.} The final policy is used in scenarios where data needs to be protected  at all times except when the device is unlocked. This functionality is achieved non-cryptographically by wiping the ClassA keys from memory 10 seconds after a device is locked. In iOS this is considered the strongest level of protection \cite{ioswhitepaper}. Note that from a cryptographic point of view, this policy provides the same key protection as the ClassC policy and the private key of the ClassB policy. The functional differences\footnote{A ClassA key will be cleared by the SEP OS 10s after the device is locked. A ClassC is only cleared when the device is powered off. This means that the $authToken$ key is used multiple times in the case of the ClassA keys and only once in the case of the ClassC keys.} are captured by the usage parameter of the policy. The policy is described as:
\begin{multline*}
    \footnotesize
    clsA = [\text{\textit{AES-256-GCM}}, \phi_{clsD}, k_{master\_key},
    \\FileProtectionComplete]
\end{multline*}

\smallskip\noindent\emph{Other keys.} In addition to the above-mentioned class keys, iOS uses three more keys: (1) the key $k_{keybag}$  secures the keybag database containing the wrapped $ClsA-C$ keys stored in the user partition, (2) the key $k_{metadata}$ encrypts the file system's metadata information (e.g. file names), and (3) the key $k_{effaceable}$ wraps both 1,2 and ClsD. The first two of these keys are always stored and managed together with $k_{clsD}$. As such, for simplicity, we refer to this group of keys ($k_{keybag}$, $k_{metadata}$, $k_{clsD}$) as  $k_{clsD}$ and make the distinction where necessary. Finally, $k_{effaceable}$ is outside the scope of this work as it provides no additional data confidentiality, and is instead designed to aid with quick erase. Disposing of this key renders all files cryptographically inaccessible.


\section{Overview}

When analysing key management for MFBE, first we want to distinguish between the following two states of device: the \textit{provisioned state}, when cryptographic keys have been initialised on the device, and the \textit{not provisioned state}, when key material is missing from the device. A device will always be in the provisioned state with the exception of right before being powered on for the first time or right after it has been restored to default settings. While a device in this state it will follow the \textit{first boot sequence}. In every other case the device will be in a provisioned state and will follow the \textit{regular boot sequence}.

As part of the \textit{regular boot sequence} there are three core stages for both Android and iOS: device unlock, first user-unlock, and regular file access after first user-unlock. In Figure \ref{fig:mfbe} the regular boot sequences for Android and iOS are presented using the notation presented earlier.

The main difference between the \textit{first boot sequence} and the \textit{regular boot sequence} is that the former includes a key generation stage which  precedes all the stages of the latter. The key generation algorithm is detailed in Section \ref{sec:keymanagement}. However it is worth mentioning that while the iPhone's SEP has a random number generator that is used to directly generate the key material inside the secure enclave, in Android the key material is generated in the non-trusted environment and then exported to the enclave via system services that run in the user space. In both systems, after the initial key generation the boot process continues with the \textit{regular boot sequence}.

        \begin{algorithm}[H]
            \footnotesize
        	\caption{$\WrapKey^\sigma$ in Android}
        	\label{alg:androidwrapkey}
        	
        	\SetKwProg{function}{function}{}{}
            \KwIn{$k, w, pol$}
            \KwOut{$\phi, \overline{pol}$}
            \Constants{$\sigma=\{\sigma\text{-}key,\sigma\text{-}ciph\}$}
        	\BlankLine
        	\BlankLine
        	
            \function{$\WrapKey(k, w, pol)$}{
                $ciph \xleftarrow{} pol.cipher$\\
                $\overline{pol} \xleftarrow{} pol$\\
                    
                \BlankLine
                \tcc{Encrypt class keys $k$ with wrap key and policy cipher.}
                
                \text{\texttt{/* DE class has no user token.\hspace{38pt} */}}
                
                \uIf {\hspace{5pt}pol.usage = DeviceDataAfterBoot \\\hspace{3pt}\text{\textbf{or}} pol.usage = UserDataAfterBoot
                \\\text{\texttt{/* CE class verifies the user token.\hspace{10pt} */}} 
                \\\hspace{3pt}\text{\textbf{or}} $(pol.usage = UserDataAfterAuth$ \\\hspace{15pt}\text{\textbf{and}} VerifyToken(pol.authToken))\\\hspace{-10pt}}
                {
                    $\phi \xleftarrow{} ENC^{ciph}(k, w)$\\
                    
                    \BlankLine
                    \tcc{Encrypt wrap key w and bind it to the policy.}
                    $\overline{pol}.wrapkey \xleftarrow{} ENC^{\sigma\text{-}ciph}(w, \sigma\text{-}key)$\\\label{alg:awrapkeyprot}
                    \Return $\{\phi, \overline{pol}\}$
                    
                } \Else {
                    \Return $\bot$
                }
            }
        \end{algorithm}
        \begin{algorithm}[H]
            \footnotesize
            \caption{$\WrapKey^\sigma$ in iOS}
        	\label{alg:ioswrapkey}
        
        	\SetKwProg{function}{function}{}{}
            \KwIn{$k, w, pol$}
            \KwOut{$\phi, \overline{pol}$}
            \Constants{$\sigma=\{\sigma\text{-}key,\sigma\text{-}ciph\}$}
        	\BlankLine
            \function{$\WrapKey(k,w,pol)$}{
                $ciph \xleftarrow{} pol.cipher$\\
                $\overline{pol} \xleftarrow{} pol$\\
            
                \uIf{$pol.usage = FileProtectionNone $}
                {
                    \tcc{ClassD i.e. FileProtectionNone}
                    $\phi' \xleftarrow{} k$\\
                }
                \Else{
                    \BlankLine
                    \uIf{$pol.authToken = null$}{
                        \tcc{Encrypt class $\text{B}_{\text{pub}}$ key $k$ with wrap key $w$ and policy cipher.}
                        $\phi' \xleftarrow{} ENC^{ciph}(k, w)$\\
                    }\Else{
                        \tcc{Encrypt class A, $\text{B}_{\text{prv}}$ and C keys $k$ with wrap key $w$, policy cipher and user password derived key $k_{master\_key}$.}
                        $k_{master\_key} \xleftarrow{} pol.authToken$\\
                        $\phi'' \xleftarrow{} ENC^{ciph}(k,k_{master\_key})$\\
                        $\phi' \xleftarrow{} ENC^{ciph}(\phi'', w)$
                    }
                    
                    \BlankLine
                    \tcc{Encrypt the wrapping key with the SEP's key.}
                    
                    $\overline{pol}.wrapkey \xleftarrow{} ENC^{\sigma\text{-}ciph}(w,\sigma\text{-}key)$
                }
                
                \BlankLine
                \tcc{Encrypt wrapped keys $\phi'$ with the hw. key.}
                    
                $\phi \xleftarrow{} ENC^{\sigma\text{-}ciph}(\phi', \sigma\text{-}key)$\label{ios_elayer0}
                
                \Return $\{\phi, \overline{pol}\}$
            }
        \end{algorithm}

\smallskip\noindent\textit{Device unlock} is the first stage of the boot process during which the device will decrypt the system partitions/files. There are some important differences between Android and iOS in this stage. As mentioned before, in Android key management is performed through services that run in the user space. A consequence of this is that the boot process needs to reach the stage where user space programs can be executed. This in turn means that the main system partition has to be unencrypted so that important system components, such as the kernel, can be loaded. Once the kernel and relevant key management services are up and running, the main encrypted device keys, $k_{DE_{\{u,s\}}}$ will be loaded from disk, decrypted inside the secure enclave, exported back to the user space key management service and finally loaded into the Android kernel memory space. Once the keys are available to the kernel relevant locations on the data partition will be decrypted to allow basic usage of the device (e.g. WiFi configuration details, phone and messaging apps).

In iOS all encryption, decryption and key management is performed inside the SEP. As such, in iOS the system partition can also be encrypted. In the beginning stages of the boot process the iOS kernel will load the encrypted class D key, $\phi_{clsD}$, which protects the whole filesystem, from the system partition into the SEP to be decrypted and enabled for use. Once the system and user partitions are decrypted by $k_{clsD}$ the ciphertexts for the remaining class keys are loaded from the user partition into the SEP.

We consider the device boot stage complete when all the keys restricting access to system related functionality have been loaded.

\smallskip\noindent\textit{User unlock} is the second stage of the boot process. During this stage the device will decrypt and load the user specific keys. Access to these keys will be conditioned on a user specific input, such as a password or biometric data, which we refer to as \textit{Passcode}. As detailed in Section \ref{sec:authtokens}, although the passcode is handled differently between Android and iOS,  the functionality remains the same in the two platforms. Once the passcode is verified, the key loading process follows a platform specific process similar to the one described in the \textit{Device unlock} stage.

\smallskip\noindent\textit{File encryption/decryption unlock} is the final stage of the boot process and represents the regular use of the device. In both Android and iOS encryption and decryption are performed transparently by algorithms running in the kernel or SEP respectively. Android uses the Linux ext4 filesystem encryption and decryption algorithm which we discuss in more detail in Section \ref{sec:linuxenc}. We were unable to find similar information related to the algorithms used in iOS. 

\section{Key management in mobile FBE schemes}
\label{sec:keymanagement}

In this section we give the generic algorithms needed to instantiate file-based encryption in Android and iOS. We show that our models capture the complete functionality of the actual implementations. In Fig. \ref{fig:mfbe} we show the interaction between the algorithms defined in this section, for the device boot, unlock and file decryption stages.

\medskip\noindent\emph{The $KGen$ algorithm.} We wanted to allow flexibility in our definition of \schemeacr{} such that our syntax and proofs could easily adapt both to existing schemes and to future schemes that aim to provide file-based encryption. To this end we have defined the KGen as the probabilistic key generation algorithm which is called whenever a random key is needed by any other \schemeacr{} algorithm. This algorithm only takes as input a security parameter and a key format. The output is a random sting that can be used as a key for that specific format.

\medskip\noindent\emph{The $\WrapKey$ algorithm.} This algorithm represents the first \schemeacr{} specific algorithm. It uses the raw keys generated by $KGen$ and protects them using one or more layers of encryption, depending on the type of key, i.e. the key policy. The $\WrapKey$ takes as input the key to be protected $k$, a wrapping key $w$, and a usage policy $pol$. The purpose of the $\WrapKey$ algorithm is to cryptographically bind input values $k$, and $pol$ to the trusted execution environment through the wrap key $w$. We model the TEE through the constants $\sigma=\{\sigma\text{-}key,\sigma\text{-}ciph\}$ representing the key stored exclusively in the TEE and the cipher associated with that key, respectively. An important aspect to notice is that, if no wrapping key is given as input (i.e. $w$ is $null$) the input key $k$ will be protected by the TEE (using the key $\sigma\text{-}key$ and the cipher $\sigma\text{-}ciph$).

We give two $\WrapKey$ algorithms, one for Android (Algorithm \ref{alg:androidwrapkey}) and one for iOS (Algorithm \ref{alg:ioswrapkey}) to describe how the key wrapping is performed in these systems. While these algorithms are very similar to one another we would like to distinguish a few differences between them.

\smallskip\noindent\textit{Android.} Due to the hardware constraints imposed on devices that run Android (i.e. very limited memory available exclusively to the TEE) on this platform we distinguish two main types of keys: wrap keys and policy-based encryption keys. As previously stated, in Android each policy-based encryption key $k$ is protected by a hardware-backed wrapping key. However, because of the limited storage available in the TEE these wrapping keys are stored outside the TEE, encrypted with the key $\sigma\text{-}key$. In Algorithm \ref{alg:androidwrapkey} we model these wrapping keys as part of the policy. The hardware-backing operation is shown on line 11.

Another important implementation decision in Android is that the $\WrapKey$ algorithm which protects device encryption keys (e.g. $DeviceDataAfterBoot$ policy) is identical to the one which protects credential encryption keys (e.g. $UserDataAfterAuth$ policy). The only difference between the two being that the $UserDataAfterAuth$ policy requires a valid authentication token before the wrapping (or unwrapping) is performed as shown on line 9.

As such, for a wrapping key $w$ and a encryption key $k$, Algorithm \ref{alg:androidwrapkey} works as follows. First the wrapping cipher is retrieved from the policy (line 2). Then the policy is checked for usage (lines 5-8). If the $pol.usage$ field contains a supported key usage policy then the encryption key $k$ is then encrypted with $w$ using the cipher $ciph$ specified in the policy (as shown on line 11). Finally, the wrap key is encrypted inside the TEE and added to the policy, thus binding key $k$ to the TEE through the policy pol, which now contains the ciphertext of the wrapped key (line 12).

\smallskip\noindent\textit{iOS.} The $\WrapKey$ algorithm for iOS is presented in Algorithm \ref{alg:ioswrapkey}. This algorithm is similar in functionality to the Android WrapKey. However, iOS does not use wrapping keys, instead this functionality is achieved using multi-layered encryption as follows.

Currently, iOS distinguishes four policies for its policy-based encryption keys: ClassA, ClassB, ClassC, ClassD. Each policy has an associated encryption key. These keys are used both for wrapping other, more secure policy keys, as well as providing the file based encryption functionality under the current policy. The keys can be grouped into three security levels based on the number of encryption layers used to wrap them.

Similarly to Android, iOS has a device unique key and cipher $\sigma$ available only to the TEE. The outer-most encryption layer uses this key to protect all the iOS keys (Algorithm \ref{alg:ioswrapkey}, line 3) and bind them to the secure hardware element.

The encryption key for the ClassD policy, $k_{clsD}$ is the least-protected key in iOS, being encrypted only with the TEE stored, device key $\sigma\text{-}key$. The $k_{clsD}$ key is used both for file encryption to provide the iOS defined \textit{FileProtectionNone} security level as well as for wrapping other policy keys. As such ClassD encrypts all other policy keys in iOS (Algorithm \ref{alg:ioswrapkey}, line 5).

The public key for the ClassB policy, $k_{clsB_{pub}}$ is the next least-protected key, being encrypted with $k_{clsD}$ and $\sigma\text{-}key$. This key is only used for file encryption, to provide write only capability to apps under the \textit{FileProtectionCompeteUnlessOpen} security policy usage (Algorithm \ref{alg:ioswrapkey}, lines 8).

Finally, keys used by ClassA and ClassC policies ($k_{clsA}$ and $k_{clsC}$), and the private key of the ClassB policy ($k_{clsB_{prv}}$) are protected with three layers of encryption. In addition to the $k_{clsD}$ and $\sigma\text{-}key$ outer layers, a third inner encryption layer is provided using a user authentication PIN/password/biometric derived key, $k_{master\_key}$. This use of this key effectively ensures that $k_{clsA}$, $k_{clsC}$ and $k_{clsB_{prv}}$ cannot be accessed unless the user is able to successfully authenticate and produce the user authentication derived key (\ref{alg:ioswrapkey}, lines 11-12). In order to more accurately compare the Android and iOS algorithms we are modelling $k_{master\_key}$ as an authentication token using our policy syntax.


\medskip\noindent\emph{The $\UnwrapKey$ algorithm.} $\UnwrapKey$ performs the opposite operation to that of $\WrapKey$. That is, given a wrapped key $\sigma$ and the policy object associated with that key $\overline{pol}$, then return the decrypted class key $k$ (for installation into the file manager's keyring). All platforms and class keys that we consider are hardware bound, hence the algorithm also requires access to the secure hardware (TEE or enclave).

A key element of $\UnwrapKey$ is to verify the context within which it is executed. This is to ensure that keys cannot be arbitrarily unwrapped without concern for the state of the device. The policies for each platform state their usage scenarios, and these are checked when executing $\UnwrapKey$. In general, the stricter policies require valid $authTokens$ (see section \ref{sec:authtokens}) to successfully unwrap the class key. As before, we will now differentiate the two implementations of $\UnwrapKey$ on each considered platform.

\smallskip\noindent\textit{Android.} In Android, each policy contains it's own hardware protected wrapping key. In Algorithm \ref{alg:androidunwrapkey}, we can see that this is decrypted by the TEE on line 3. If the DE policies are being initiated, then this wrap key is then directly used to decrypt the class key (line 9). Otherwise, when the more secure CE policy is in use, the system requires a successful generation and verification of the $authToken$ to proceed (line 7). Although authTokens are not cryptographically used to protect the CE key, the TEE will not proceed with the key decryption until a $authToken$ has been provided and verified. 

\smallskip\noindent\textit{iOS.} Unwrapping keys in iOS consists of iteratively removing layers of encryption. Each call to the function will first begin by removing the hardware encryption layer, which all supported class keys are protected with (line 3). As described in the previous section, each class key has varying encryption layers and dependencies. The logic which defines how to decrypt each key is defined in lines 4-13. Notably, to remove the final encryption layer on the most secure class keys (A, Bprv, and C), the iOS version of a valid authToken is required (i.e. the $k_{master\_key}$). In contrast to Android, this $k_{master\_key}$ is directly used to encrypt the keys belonging to the relevant classes. 

\medskip\noindent\emph{The $\ProvisionKey$ \&  $\EvictKey$ algorithms (Appendix \ref{app:algorithms} alg. \ref{alg:provisionkey} \& \ref{alg:evictkey}).} In Android and iOS keys are not accessed directly, but through data structures called \textit{key handles} that are  similar to memory pointers. Use of these handles allows the actual key plaintext to be stored securely, for example in the TEE's memory while, at the same time, enabling programs running outside of this context, (e.g. user-land programs), to use the keys for cryptographic operations, without actually having access to the key itself. We capture this functionality through the $\ProvisionKey$ algorithm. This algorithm simply takes as input a plaintext key and returns a key handle $kid$ which can safely be exported and used outside the hardware context which performs the actual cryptographic operations. The $\ProvisionKey$ algorithm plays an important role in the security of the scheme, as running it in the wrong context could lead to the key plaintext being stored in an unsafe memory location or leaked.

The $\EvictKey$ algorithm is the algorithm used to clear the plaintext of a key from memory using the key's handle, $kid$. Key should be an operation that can be safely run in any context as it does not have direct access to any cartographic key material.

\begin{algorithm}[t]
    \footnotesize
    \caption{UnwrapKey$^\sigma$ in Android}
	\label{alg:androidunwrapkey}

	\SetKwProg{function}{function}{}{}
	\KwIn{$\phi, \overline{pol}$}
    \KwOut{$k$}
    \Constants{$\sigma=\{\sigma\text{-}key,\sigma\text{-}ciph\}$}
    
 	\BlankLine
    \function{$UnwrapKey(\phi, \overline{pol})$}{
        $ciph \xleftarrow{} \overline{pol}.cipher$\\
        $w \xleftarrow{} DEC^{\sigma\text{-}ciph}(\overline{pol}.wrapkey, \sigma\text{-}key)$\\
    
        \uIf {\hspace{5pt}pol.usage = DeviceDataAfterBoot \\\hspace{3pt}\text{\textbf{or}} pol.usage = UserDataAfterBoot
        \\\hspace{3pt}\text{\textbf{or}} $(pol.usage = UserDataAfterAuth$
        \\\hspace{15pt}\text{\textbf{and}} VerifyToken(pol.authToken))
        \\\hspace{-10pt}}
        {
            $k \xleftarrow{} DEC^{ciph}(\phi, w)$\\
            return $k$
        } \Else {
            return $\bot$
        }
    }
\end{algorithm}

\section{File Content Encryption in Mobile FBE Schemes}
\label{sec:ext4enc}

In this section we propose a formal description of the file encryption scheme used  by the EXT4 filesystem as it has been implemented in Android and *NIX systems. We have not been able to access the source code of the software produced by Apple (i.e. all Apple software is closed source). However, we speculate that the iOS implementation of FBE is conceptually similar to the EXT4 implementation.

\subsection{Android EXT 4 Case Study}
\label{sec:linuxenc}

The EXT4 encryption is a FBE scheme originally developed by Google \cite{androidfbe} for use with the Android 7.0 and higher. It became part of the mainline kernel in version 4.0 under the name of Filesystem-level encryption (fscrypt) and has suffered some adaptations. Google continues to maintains the corresponding User space tool (also named fscrypt) for use with the kernel API discussed in the following.\footnote{https://github.com/google/fscrypt}

\begin{algorithm}[t]
    \footnotesize
    \caption{UnwrapKey$^\sigma$ in iOS}
	\label{alg:iosunwrapkey}
	
	\SetKwProg{function}{function}{}{}
	\KwIn{$\phi, \overline{pol}$}
    \KwOut{$k$}
    \Constants{$\sigma=\{\sigma\text{-}key,\sigma\text{-}ciph\}$}
 	\BlankLine
    \function{$UnwrapKey(\phi, \overline{pol})$}{
        $ciph \xleftarrow{} \overline{pol}.cipher$\\
        \tcc{Remove the hardware encryption layer from $\phi$.}
        $k' \xleftarrow{} DEC^{\sigma\text{-}ciph}(\phi, \sigma\text{-}key)$\\
        
        \uIf{$pol.usage = FileProtectionNone $}{
            \tcc{FileProtectionNone i.e. ClassD}
            
            $k \xleftarrow{} k'$\\
        
        } \Else{
            \tcc{Decrypt wrap key $w$.}
            $w \xleftarrow{} DEC^{\sigma\text{-}ciph}(\overline{pol}.wrapkey, \sigma\text{-}key)$\\
            
            \tcc{Decrypt class $\text{B}_{\text{pub}}$ and the 1st layer for class A, $\text{B}_{\text{prv}}$ or class C keys with wrap key $w$ and policy cipher.}
            $k'' \xleftarrow{} DEC^{ciph}(k', w)$\\
            
            \BlankLine
            \uIf{$pol.authToken = null$}{
                $k \xleftarrow{} k''$
            } \Else {
                \tcc{Decrypt class A, $\text{B}_{\text{prv}}$ or C keys $k$ with user key.}
                
                $k_{master\_key} \xleftarrow{} pol.authToken$\\
                $k \xleftarrow{} DEC^{ciph}(k'',k_{master\_key})$\\
            }
        }
        
	    \BlankLine
		\Return $k$
    }
\end{algorithm}

There are currently three main modes in which fscrypt can function on Android. First  is the original scheme developed by Google for Android 7.0. This scheme is currently summarised under the \textit{v1 policy} in the official documentation \cite{EXT4-whitepaper} and suffered from several limitations as follows. Even though per-file encryption keys were derived from a master key (such as $k_{CE}$), compromising one of these could result in compromising the master key from which the per-file key was derived from. This is mainly due to the odd use of a publicly known random value, $n$ as the key input to the AES cipher used for key derivation. This is captured in Algorithm \ref{alg:seckey}, line 3 and is described in more detail in the next section. Additional limitations to v1 were also the lack of verification for the master key usage which would allow malicious users to gain read only access to other user's encrypted files, and also there were difficulties in removing encryption keys from memory in order to limit access to protected files \cite{EXT4-whitepaper}. Addressing these issues has resulted in an enhanced fscrypt and a new key derivation function, summarised under the \textit{v2 policy}.

Finally, starting with Android 10 fscrypt started supporting a new cipher, Adiantum, developed by Google \cite{crowley2018adiantum} and aimed at performance.\footnote{https://security.googleblog.com/2019/02/introducing-adiantum-encryption-for.html} With the addition of this new cipher a new mode of operation for fscrypt was also added, Adiantum.Direct\_Key which no longer uses a KDF. The Adiantum cipher however can also be used with v1 and v2 policies.

In the following we propose several algorithms which summarise and simplify the way in which fscrypt is implemented in Android (and the Linux kernel). These algorithms also allow us to directly compare the security and features of the various modes supported currently and in the future by fscrypt.


The EXT4 filesystem uses the notion of \textit{nodes} to generically describe its components: files, directories and symbolic links. In our pseudo-code we have used the same notation to describe the sequence of operation needed to perform FBE operations. To simplify notation, we define an EXT4 node as a tuple $node=(name, \allowbreak meta, \allowbreak type, \allowbreak content, \allowbreak children)$ where:

\begin{itemize}[leftmargin=0.4cm]
    \item[--] $node.name$: is the file or directory name of the node.
    \item[--] $node.meta$: is a storage container associated with the node different from the main content of the file. It stores metadata information such as permissions, data access information, etc., but also the \textit{xattr} attribute which stores the unique cryptographic nonce of the node. 
    \item[--] $node.type$: represents the type of the node, i.e. file or directory.
    \item[--] $node.content$: in case $node.type$ is a \textit{file} the \textit{content} represents the main storage container for the node. This mainly refers data stored inside files.
    \item[--] $node.children$: in case $node.type$ is a \textit{directory} the $node.children$ will be a set of all the $node.names$ contained inside that node, i.e. the names of the file and sub-directories contained in the specified node.
\end{itemize}

\begin{algorithm}[t]
    \footnotesize
	\SetKwProg{function}{function}{}{}
	\KwIn{$\textit{f}node,kid,mode$}
	\KwOut{$enode$}
	\BlankLine
	\function{$\mathsf{\Encrypt}(\textit{f}node,kid)$}{
		$k_m  \xleftarrow{} {} GetKeyFromKernelKeyring(kid)$\\
		\If {$k$}{
		    $n \xleftarrow{} \fnode.meta$\\
		    \If {$n = null$}{
                \tcc{New fnode.}
                $n \xleftarrow{R} \{0,1\}^{128}$\\
            }
            $\enode.meta \xleftarrow{} n$\\
            \BlankLine
            
            \tcc{Filename/metadata is protected by directory sec context}
            \If {$\textit{f}node.type = directory$}{
                $k_e \xleftarrow{} KDF(k_m, n, mode)$\\
                $enode.children \xleftarrow{} ENC^\fileciph(\textit{f}node.children, k_e)$\\
            }
            
            \BlankLine
            
            \tcc{File contents is protected by file sec context}
		    \ElseIf {$\textit{f}node.type = file$}{
                $k_e \xleftarrow{} KDF(k_m, n, mode)$\\
                $enode.content \xleftarrow{} ENC^\fileciph(\textit{f}node.content, k_e)$\\
            }
            \Return $enode$
		} \Else {
            \Return $\bot$
		}
	}
	\caption{File encryption in EXT4 filesystems (Android \& *NIX systems)}
	\label{alg:encryptfbe}
\end{algorithm}

Additionally, to facilitate comparison between the supported FBE modes of operation we also define the encryption keys as $k=(mode, \allowbreak val)$ where $k.mode$ represents the fscrypt mode in which the key is used and $k.val$ is the actual binary value of the key.

In order to perform encrypt (or decrypt) operations a master key is required. There can be any number of master keys in use simultaneously on a system. In Android these keys are generated and made available to the kernel as described in Section \ref{sec:keymanagement}. In regular *NIX systems these keys need to be made available to the kernel by the \textit{User space} \cite{EXT4-whitepaper}. Once a key is loaded into the kernel a unique key id, $kid$, is generated for use in \textit{User space}. All encrypt and decrypt operations will be managed transparently by the kernel, and keys will be specified through their ids.

EXT4 encryption does not use the master keys directly, instead, it uses individual keys for each $node.name$ and each $node.content$. These keys are derived from the master key based on the fscrypt policy used and the 128bit unique random nonces associated with each node. The policy will define the ciphers used for encryption (and decryption), including the generation and use of IV values, and the key derivation function (KDF) to be used for generating individual per-file keys.

All policies employ block-chaining modes for disk encryption when encrypting the $node.content$ and thus provide similar level of security. These modes make use of hardware dependent values such as the logical block number (lbn) of the storage device to generate unique IV values. This however is not the case when encrypting $node.names$, and thus the policy plays a more significant role in how much detail is revealed about these.

\begin{algorithm}[t]
    \footnotesize
	\SetKwProg{function}{function}{}{}
	\KwIn{$k,n,mode$}
	\KwOut{$k_e$}
	\BlankLine
	\function{$\mathsf{KDF}(k,n,mode)$}{
        \uIf {$mode = \mathsf{v_1}$}{
            \tcc{Original mode}
            ${k_e}.\mathsf{val} \xleftarrow{} AES\text{-}128\text{-}ECB(k,n) $\\
        }
        \uElseIf{$mode = \mathsf{Adiantum.DIRECT\_KEY}$}{
            ${k_e}.\mathsf{val} \xleftarrow{} k$\\
        }
        \uElseIf {$mode = \mathsf{IV\_INO\_LBLK\_64}$}{
            ${k_e}.\mathsf{val} \xleftarrow{} \mathsf{HKDF\text{-}SHA512}(k, mode \| \mathsf{<uuid\footnotemark>})$\\
        }
        \Else{
            \tcc{Default: per-file keys}
            $cb \xleftarrow{R} \{0,1\}^8 $\\
            ${k_e}.\mathsf{val} \xleftarrow{} \mathsf{HKDF\text{-}SHA512}(k,\mathsf{"fscrypt"} \| cb \| n\footnotemark) $\\
        }
        ${k_e}.mode \xleftarrow{} mode$\\ 
        \Return $k_e$
	}
	\caption{Key derivation (Android \& *NIX systems)}
	\label{alg:seckey}
\end{algorithm}
\footnotetext[6]{UUID: Universally Unique Identifier used to identify disk partitions.}
\footnotetext{Literal string "fscrypt" concatenated with a context byte ($cb$) and $n$.}

\begin{algorithm}[t]
    \small
	\SetKwProg{function}{function}{}{}
	
	\KwIn{$content,k$}
	\KwOut{$econtent$}
	\BlankLine
	\function{$ENC^{\fileciph}(content, k)$}{
        \For {$cblock \in content$}{
        	$\mathsf{iv} \xleftarrow{} \mathsf{GetIV}(cblock,k.mode)$\\
            $eblock \xleftarrow{} CIPHER(cblock, k.\mathsf{val},\mathsf{iv})$\\
            $econtent \xleftarrow{} econtent \| eblock$\\
        }
    }
	\caption{Node content encrypt (Android \& *NIX systems)}
	\label{alg:secenc}
\end{algorithm}

\medskip\noindent\textit{1) Fscrypt key derivation methods}

\noindent\textit{The v1 policy.} In this policy the individual per-file keys are derived from a master key using a KDF based on AES-256-ECB by simply encrypting the master key with the node's random 128bit nonce as shown in Algorithm \ref{alg:seckey} line 3. This method was chosen to reduce the size of the cryptographic metadata stored on disk. It is noted however by the authors that this requires the derived key to be "equally as hard to compromise" as the original master key and is currently considered to be a security limitation \cite{EXT4-whitepaper}.

\medskip\noindent\textit{The v2 policy.} This policy replaces the AES-256-ECB based KDF with a HKDF-SHA512 for generating per-file keys. In addition to being irreversible, the new derivation method also takes as input an application-specific string in addition to the master key. These modification successfully mitigate the problems related to master key recovery from per-file encryption keys as well as key reuse between applications (Algorithm \ref{alg:seckey} lines 9-10). The v2 policy also supports generating keys for use with encryption hardware compliant with the UFS or eMMC standards, under the IV\_INO\_LBLK\_64 sub-policy (Algorithm \ref{alg:seckey} line 6-7). This sub-policy also enforces a specific form for the IV as detailed below.

\medskip\noindent\textit{Adiantum.DIRECT\_KEY.} This is not a policy in itself rather the absence of one. In \textit{Adiantum.DIRECT\_KEY} the master key is directly used to encrypt $node.name$ and $node.content$ for all the nodes in the file system. This mode however is restricted to the Adiantum cipher as it relies on the cipher's ability to make use long IVs as described below.

\medskip\noindent\textit{2) Encryption}

\noindent Currently, Fscrypt supports AES-256-XTS,  AES-128-CBC and Adiantum for $node.contents$ encryption and AES-256-CTS-CBC, AES-128-CTS-CBC and Adiantum for $node.name$ encryption. Each of these ciphers will use different methods for generating valid IV's based on the key derivation method, and whether they are used for encrypting $node.names$ or $node.content$.

\medskip\noindent\textit{Node's content encryption.} In the default configuration for encrypting the $\allowbreak node.content$ under both \textbf{v1} and \textbf{v2} key derivation methods the IV is initialised with the logical block number of the block residing on the storage device. This IV can be used directly with AES-XTS and Adiantum (in non DIRECT\_KEY configuration). AES-CBC requires a random IV to achieve the desired security properties, as such this is achieved by encrypting the \textit{lbs} with a SHA256 hash of the per-file encryption key (see Algorithm \ref{alg:seciv}, lines 7-9).

An exception also exists for the IV\_INO\_LBLK\_64 sub-policy which describes the generation of a 64bit IV from a 32bit truncated \textit{lbn} and another 32bits limited value representing the nodes \textit{innode number} \footnote{a uniquely existing number for all the nodes on the EXT4 partition} (Algorithm \ref{alg:seciv}, lines 10-11). Under \textit{Adiantum.DIRECT\_KEY} the IV is obtained by concatenating the full 64bit \textit{lbn} with the node's random nonce as shown in Algorithm \ref{alg:seciv}, lines 12-13.

\medskip\noindent\textit{Node's name encryption.} Like most filesystems EXT4 is structured in a tree-like format. This enables specific filesystem functionality, such as managing permissions, browsing contents, searching, deleting, etc., to be efficient. The \textit{fscrypt} has been implemented such that encrypted EXT4 filesystems will continue to have all the functionality of non-encrypted EXT4 filesystems.

As such, to support some of these operations, node names are encrypted using keys associated to the parent node, e.g. the name of a file is encrypted with the key corresponding to the directory where is it stored. However, node names (and other metadata associated with nodes) are stored separately from the node's contents, and device blocks are often shared when storing, thus \textit{lbn} values cannot be used as IV values. Similarly, block chaining modes that rely on the \textit{lbn} XTS are also unsuitable.

In order to address this fscrypt restricts the ciphers and modes to AES-CTS-CBC and Adiantum when encrypting node names. In the default configuration an all-zero IV is used (Algorithm \ref{alg:seciv}, line 6). Given that the key is also shared between all children inside a node this results in encryption deterministic which leaks information about the names. This limitation is mitigated when Adiantum.DIRECT\_KEY and IV\_INO\_LBLK\_64 policies are used as the IV in these cases are less dependent on the \textit{lbn}. We detail this in Algorithm \ref{alg:seciv}. 

\medskip\noindent The cryptographic algorithms used for the encryption and decryption of file names in a directory (i.e. node name) are very similar to the ones used for encrypting blocks of content in a file (i.e. node contents). We model both in Algorithms \ref{alg:secenc} and \ref{alg:secdec} respectively, with the added clarification that in the former case the iteration is done on $node.names$ whereas in the latter on $\textit{f}sblocks$.

\begin{algorithm}[t]
    \footnotesize
	\SetKwProg{function}{function}{}{}
	\KwIn{$\textit{f}sblock,mode$}
	\KwOut{$\mathsf{iv}$}
	\BlankLine
	\function{$\mathsf{GetIV}(\textit{f}sblock,mode)$}{
	    $\textit{f}node \xleftarrow{} GetInodeOf(\textit{f}sblock)$\\
	    
	    \If{$\textit{f}node.type = file$}{
            $lbn \xleftarrow{} \textit{f}sblock.lbn$\\
        }
        \Else { \tcc{Directory}
            $lbn \xleftarrow{} \{0\}^{128}$\\
        }
        
        \BlankLine
        \BlankLine
        
        \uIf{$mode = \mathsf{AES\text{-}128\text{-}CBC}$}{
            $k_{\mathsf{iv}} \xleftarrow{} \mathsf{SHA256}(k)$\\
            $\mathsf{iv} \xleftarrow{} \mathsf{AES\text{-}256\text{-}ECB}(lbn,k_{\mathsf{iv}})$\\
        }
        \uElseIf {$mode = \mathsf{IV\_INO\_LBLK\_64}$}{
            $\mathsf{iv} \xleftarrow{} lbn_{len=32b} \| \textit{f}node.InodeNumber_{len=32}$\\
        }
        \uElseIf{$mode = \mathsf{Adiantum.DIRECT\_KEY}$}{
            $\mathsf{iv} \xleftarrow{} lbn_{len=64b} \| \textit{f}node.meta_{len=128b}$\\
        }
        \Else{
            \tcc{Default mode}
            $\mathsf{iv} \xleftarrow{} lbn$
        }
        \Return $\mathsf{iv}$
	}
	\caption{Get IV (Android \& *NIX systems)}
	\label{alg:seciv}
\end{algorithm}

\section{Security analysis models}
In this section we consider several adversarial models that affect the security of FBE in mobile devices

\begin{figure*}[t!]
	\hspace{-20pt}
	\framebox[1.11\textwidth][t]{
		\small
		\hspace{8pt}
		\begin{tabular}{l}
		    \hspace{-15pt}
		    
    		\begin{tabular}{p{240pt}  p{115pt} p{120pt}}
        		\begin{tabular}{l}
        			$Exp_{\mathcal{A}}^{\schemesymb,b}(\lambda)$\\[5pt]\hline
        			$ \sigma \xleftarrow{R} \{0,1\}^*$ \\[2pt]
        			$ K:\{(k,kid)\vert(\forall(k',kid')\in K)[kid\neq kid']\}$\\
        			$ K \xleftarrow{} \emptyset$\\
        			$ (state,\textit{f}node_0, \textit{f}node_1,kid^\star) \xleftarrow{} \mathcal{A}_{1}^{\mathcal{O}^{wrap},\mathcal{O}^{unwrap},\mathcal{O}^{reveal}, \calO^{enc}}(1^\lambda)$\\
        			$enode_b\xleftarrow{} \mathsf{Encrypt}(kid^\star,\textit{f}node_b)$\\
        			$ b' \xleftarrow{} \mathcal{A}_{2}^{\mathcal{O}^{wrap},\mathcal{O}^{unwrap}, \mathcal{O}^{reveal}, \calO^{enc},\mathcal{O}^{dec}}(state,enode_b)$\\
        			
        			$\mathbf{return}~ b'$
        		\end{tabular}
        	\end{tabular}
            
            \\\\
            
            \hspace{-15pt}
            \begin{tabular}{p{0.4\textwidth} p{0.4\textwidth}  p{0.4\textwidth} p{0.35\textwidth}}
            
                \begin{tabular}{l}
        			Oracle $\mathcal{O}^{wrap}(pol)$\\\hline
        			$ k,w \xleftarrow{} \KGen(1^\lambda)$ \\[2pt]
        	        $ (\phi,\overline{pol}) \xleftarrow{} \mathsf{WrapKey}^\sigma(k,w,pol)$ \\
        			$\mathbf{return}~ (\phi,\overline{pol})$\\
        			\\
        			\\
        			\\
        			\\
    		    \end{tabular}
    		    
    		    &
    		    
    		    \begin{tabular}{l}
        			Oracle $\mathcal{O}^{unwrap}(\phi,\overline{pol})$\\[-1pt]\hline
        	        $ k \xleftarrow{} \UnwrapKey^\sigma(\phi,\overline{pol})$ \\[2pt]
        	        $\mathbf{if}~ k ~\mathbf{then:}$\\
        	            \quad$ kid \xleftarrow{} \mathsf{ProvisionKey} (k,\overline{pol})$\\
        			    \quad$ K \xleftarrow{} K \cup \{(k, kid)\}$\\
        			    \quad$ \mathbf{return}~ kid$\\
        			$\mathbf{else:}$\\
        			    \quad$ \mathbf{return}~ \bot$
                \end{tabular}
                
                & &
                
                \\
            
                \begin{tabular}{l}
    			Oracle $\mathcal{O}^{enc}(kid, \fnode)$\\[2pt]\hline
    			$\mathbf{if}~ (\cdot,kid)\in K ~\mathbf{then:}$\\
    			    \quad$enode \xleftarrow{} \mathsf{Encrypt}(kid,\fnode)$\\
    			    \quad$\mathbf{return}~ \enode$\\
    			$\mathbf{else:}$\\
    			    \quad$ \mathbf{return}~ \bot$\\
	            \end{tabular}
                
                & 
                
                \begin{tabular}{l}
        			Oracle $\mathcal{O}^{dec}(kid,enode)$\\[2pt]\hline
        			$\mathbf{if}~ (\cdot,kid)\in K ~\mathbf{then:}$\\
        			    \quad$\textit{f}node \xleftarrow{} \mathsf{Decrypt}(kid,enode)$\\
        			    \quad$\mathbf{return}~ \textit{f}node$\\
        			$\mathbf{else:}$\\
        			    \quad$ \mathbf{return}~ \bot$\\
        	    \end{tabular}
                
                &
                
                \begin{tabular}{l}
        			Oracle $\mathcal{O}^{reveal}(kid)$\\[2pt]\hline
        			$\mathbf{if}~ (k,kid)\in K ~\mathbf{then:}$\\
        			    \\
        			    \quad$ \mathbf{return}~ k$\\
        			$\mathbf{else:}$\\
        			    \quad$ \mathbf{return}~ \bot$
                \end{tabular}
                
                &  \\
            \end{tabular}
        \end{tabular}
	}
	\label{fig:sec_exp1}
	\caption{MobileFBE security  experiment.}
\end{figure*}

\medskip\noindent\emph{Encryption model.} This adversarial model focuses on attacks against the confidentiality of files contents and filenames that are part of the underling filesystem. This adversarial model does not consider the confidentiality of the non-filename metadata such as number of files, timestamps and permissions.
This attack also does not consider timing attacks and access frequency attacks that might reveal information about how the system is used.

\medskip\noindent\emph{Key security model.} This adversarial model focuses on attacks related to key generation and key recovery.

\medskip\noindent\emph{Permissions model.} This adversarial model focuses on attacks that break the cryptographic access controls as enforced by the keys.

\begin{definition}\label{def:mfbe}
    The security of a \schemeacr{} encryption scheme is defined through a security experiment as follows (Figure \ref{fig:sec_exp1}). Let $\schemesymb{}=\allowbreak (\allowbreak\mathsf{KGen},
\allowbreak\mathsf{WrapKey}, \allowbreak\UnwrapKey,
\allowbreak\mathsf{ProvisionKey}, \allowbreak\mathsf{EvictKey}, \allowbreak\mathsf{Encrypt}, \allowbreak\mathsf{Decrypt})$ be a mobile encryption scheme with a security parameter $\lambda$ and a two stage adversary $\adv=(\adv_1,\adv_2)$. We consider $Exp_{\mathcal{A}}^{\schemesymb,b}(\lambda)$ a probabilistic experiment defined in terms of a game played between an adversary $\adv$ and a challenger $\cdv$, consisting of:

\begin{enumerate}[leftmargin=16pt]
    \item Setup. $\cdv$ initialises the hardware key $\sigma$ with a random string and creates an empty set $K$ which keeps track of associations between keys $k$ and their unique $kid$ handles as $(k,kid)$ pairs.
    \item Query. Adversary $\adv_{1}$ queries oracles $\mathcal{O}^{wrap}$, $\mathcal{O}^{unwrap}$, and $\mathcal{O}^{reveal}$.
    \item Challenge. The challenger receives from the adversary $\adv_{1}$ a  $kid^\star$, and two file nodes $\textit{f}node_0$ and $\textit{f}node_1$ , such that
$\mathcal{O}^{reveal(kid^\star)}$ has never been queried. 
            $\cdv$ runs $enode^\star \get \mathsf{Encrypt}(kid^\star, \textit{f}node_b)$ to encrypt $\textit{f}node_b$, and returns $enode^\star$.
    \item Guess. $\adv_{2}$ outputs $b'\in\{0,1\}$. In addition to the oracles $\calO^{wrap}, \calO^{unwrap},  \calO^{enc}$, $\adv_{2}$ is given access to the oracle $ \calO^{reveal}$ but cannot query $\calO^{reveal(kid^\star)}$, and the oracle $\mathcal{O}^{dec}$ but cannot query $\mathcal{O}^{dec(kid^\star, enode)}$ such that $enode$ is either the challenge ciphertext or is produced by the oracle $\mathcal{O}^{enc(kid^\star, \cdot)}$. The output of the experiment is $b'$.
\end{enumerate}

\end{definition}

We capture the adversary's abilities using formal definitions of oracles in Figure \ref{fig:sec_exp1}. Below we provide the intuition for these oracles.

\begin{description}[leftmargin=16pt]
    \item [$\mathcal{O}^{wrap}$:] Allows the adversary to create $\schemesymb{}$ keys. Receives as input an policy $pol$. For this policy the oracle generates a key $k$ and a wrapping key $w$ using the key generation algorithm $KGen(1^{\lambda})$. Then the key $k$ is encrypted with the wrapping key as specified in the $WrapKey$ algorithm. Inside $WrapKey$ $w$ is also encrypted with the hardware key $\sigma$ and is returned as part of the hardware-bound policy $\overline{pol}$. The oracle maintains a state of all the generated keys k using the set $S$. Finally, the ciphertext of $k$, $\phi$ is returned together with its bounding policy $\overline{pol}$.
    \item [$\mathcal{O}^{unwrap}$:] Allows the adversary to instantiate $\schemesymb{}$ keys generated under a specific policy. Receives as input an encrypted key $\phi$ and its associated policy $\overline{pol}$. Recovers the key using the wrap key from the policy and authenticated is using the key set $S$. If the key is valid a key handle $kid$ is returned ($\bot$ otherwise). The links between key ids and keys are maintained using the key map $K$.
    \item [$\mathcal{O}^{enc}$:] Allows the adversary to encrypt a file node $\textit{f}node$ using a key id $kid$. The oracle returns the encrypted file node $enode$ if the encryption was successful and $\bot$ otherwise. 
    \item [$\mathcal{O}^{dec}$:] Allows the adversary to decrypt a previously encrypted $enode$ using a key id $kid$. The oracle returns the decrypted file node $\textit{f}node$ if the decryption was successful and $\bot$ otherwise.
    \item [$\mathcal{O}^{reveal}$:] Allows the adversary to obtain the key associated to a valid handle $kid$.
\end{description}

\begin{definition}[Security of $\schemesymb{}$]
    A mobile file based encryption scheme $\schemeacr{}$ is secure if the advantage defined as:
    
    {\footnotesize
    $$
    Adv_{\mathcal{A}}^{\schemesymb}(\lambda)=\Big\lvert\prob{{\Exp}_{\adv}^{\schemesymb,1}( \lambda ) = 1}- \prob{{\Exp}_{\adv}^{\schemesymb,0}( \lambda ) = 1}\Big\rvert
    $$
    }
    
    \noindent is negligible in $\lambda$ for any polynomial based two stage adversary $\adv=(\adv_1,\adv_2)$.
\end{definition}

Below we shall analyse the security of the Android implementation given in Algorithm \ref{alg:androidwrapkey},  \ref{alg:androidunwrapkey}, \ref{alg:encryptfbe}, \ref{alg:decryptfbe}, with the following simplification and generalisation: 
\begin{itemize}[leftmargin=10pt]
    \item In the file encryption algorithm in Algorithm \ref{alg:encryptfbe}, the file name is encrypted with a constant iv which makes the encryption deterministic and breaks the $\mfbe$ security definition. However, as mentioned before, this is used for maintaining equality relationship between encrypted and non-encrypted nodes. Therefore, in our security model, we ignore the file name encryption part, i.e., line 7-13 in Algorithm \ref{alg:encryptfbe} and line 5-8 in Algorithm \ref{alg:decryptfbe}, and only model the file content encryption. 
    Formally, 
    \begin{itemize}[leftmargin=10pt]
        \item We model the key derivation function, i.e., $ENC^{AES-128-ECB}(k, \cdot)$ as a pseudorandom function (PRF) $F_k(\cdot)$;
        \item We write $ENC^{file\texttt{-}ciph}$ and $DEC^{file\texttt{-}ciph}$ for the cipher used for encrypting/decrypting the file content.
        \item We simplify the file node structure and only consider two fields, i.e., $node = (meta, content)$. We require that the challenge file nodes $\fnode_0$ and $\fnode_1$ satisfy  $\fnode_0.meta = \fnode_0.\allowbreak meta = null$ and $\abs{\fnode_0.\allowbreak content} = \abs{\fnode_1.\allowbreak content}$.
    \end{itemize}
    \item User authentication in $\WrapKey$ (Algorithm \ref{alg:androidwrapkey}) and  $\UnwrapKey$ (Algorithm \ref{alg:androidunwrapkey}) is not relevant to the security. For simplicity, in our security analysis we omit the user authentication part, i.e., the if branch "{\bf if} $pol.usage = UserDataAfterAuth$...".
    \item We assume the cipher $ciph$ used for encrypting the class key $ENC^{ciph}(k,w)$ is fixed. If we allow users to choose $ciph$, then we need additional mechanism to guarantee each wrap key $w$ will only be used for a certain type of cipher. This can be achieved by, for example, encrypting the cipher type with the wrap key, i.e., $ENC^{\sigmaciph}(w\|ciph, \sigmakey)$.
     \end{itemize}

\begin{theorem}
The Android $\schemesymb{}$ is secure if $ciph$ and $\sigma\text{-}ciph$ are IND-CPA and INT-CTXT secure, $F$ is a family of pseudorandom functions, and $file\text{-}ciph$ is IND-CCA secure. 
Formally, for any adversary $\A$ that attacks $\mfbe$, runs in time at most $\tau$ and makes at most $q_w$ queries to $\calO^{wrap}$, there exist adversaries $\B_1, \dots, \B_6$ such that they run in time at most $\tau$ and 
   \begin{align*}
Adv&_\A^{\mfbe}  \leq 2 Adv_{\sigmaciph, \B_1}^{CTXT} + 2 Adv_{\sigmaciph, \B_2}^{CPA} + 2 q_w Adv_{ciph, \B_3}^{CTXT} \\
& \quad + 2 q_w Adv_{ciph, \B_4}^{CPA} +  2 q_w Adv_{F, \B_5}^{PRF} + q_w  Adv_{\fileciph, \B_6}^{CCA}\\
\end{align*} 
\end{theorem}

See Appendix \ref{proof} for a proof.

\medskip\noindent\emph{Cipher modes security.}
In the following we briefly recall well-known symmetric encryption schemes and their security properties:

\begin{itemize}
    \item AES-GCM \cite{McGrewV04}: The Galois/Counter Mode of Operation (GCM) is a widely deployed authenticated encryption scheme, which is proven to be IND-CTXT and IND-CCA secure \cite{NiwaOMI15} 
    \item AES-CBC: with random IV acting as a nonce achieves IND-CPA security \cite[Chapter 13]{Smart16} 
    \item Adiantum \cite{crowley2018adiantum}:  a tweakable, variable-input-length, super-pseudorandom permutation that it is used as a fast length-preserving encryption scheme that cannot achieve IND-CCA nor IND-CPA security
    
    \item Well-known generic symmetric encryption modes that achieve IND-CTXT/CCA security  include \cite{Smart16}:
\begin{itemize}
    \item IND-CPA encryption + strong MAC (e.g., HMAC/CBC-MAC) $\rightarrow$ IND-CTXT encryption
    \item IND-CPA + IND-CTXT $\rightarrow$ IND-CCA encryption
\end{itemize}



\end{itemize}

\section{Related Work}

Prior to the adoption of File-based encryption (FBE), Full-Disk encryption (FDE) technologies were the encryption method of choice for most desktops, mobiles and other devices. FDE schemes have been extensively researched both formally, and implementations thereof. See for example Rogaway  \cite{rogaway2011evaluation} and Khati et al. \cite{khati2017full} for formal analyses, and \cite{choudary2012infiltrate} for practical implementation analyses. 

The original encrypting file system, Matt Blaze's CFS \cite{blaze1993cryptographic}, transparently encrypted data passing through a user-level NFS daemon. A variation of CFS, named Cryptfs, which leveraged virtual inodes, was presented in the late 90s  \cite{zadok1998cryptfs}. This was then extended by Halcrow in \cite{halcrow2005ecryptfs}, which years later has now been updated and adopted for use in Android as the EXT4 filesystem \cite{EXT4-whitepaper}.

Early analysis of major mobile OS vendor encryption schemes started with \cite{pandya2008iphone}, \cite{belenko2011overcoming}, \cite{bedrune2011iphone} on iOS. This works were followed by a framework for high-level security analysis proposed in \cite{mobdevencsys}. The authors subsequently presented case studies on iOS \cite{teufl2013ios} and Android \cite{teufl2014android}. Our work adopts some of the threat models from these proposals, however we take things further by providing a formal analysis.
We also update the analysis of \cite{teufl2014android} with our consideration of Androids FBE scheme, whereas before only the FDE scheme was analysed. 

Related to works analysing the Android encryption schemes are those focusing on secure key storage and management. Cooijmans et al \cite{cooijmans2014analysis} evaluate various key storage solutions in Android.
Sabt et al \cite{sabt2016breaking} carry out a cryptanalysis of Android KeyStore and formally prove the encryption scheme does not provide integrity, enabling forgery attacks. Implementation flaws discovered through reverse engineering efforts include those of \cite{beniamini2016extracting} and \cite{hay2014android}. 

Due to the overhead in software based disk-encryption there has been a push for hardware solutions in recent years \cite{muller2015systematic,lee2017fessd}. Meijer et al. \cite{meijer2018self} find many critically flawed implementations of this in SSDs. They highlight the risk posed by an OS that relies on third party encryption implementations, using the example of Microsoft's BitLocker. Apple's software and hardware is tightly coupled, and as such, this diversity of vendors and implementations is less of a concern. 

\section{Conclusion}
File-based encryption schemes have been developed independently by Apple for their iOS mobile OS and by Google for Android. The initial Android proposition was subsequently extended and added to mainline Linux kernel. However, given that these schemes originate from commercial applications their security properties have only been supported by high-level descriptions and anecdotal evidence. In this work we are bridging this gap by rigorously proving the industry claimed security properties for the analysed schemes. Furthermore, through our proposed definitions and evaluation models we hope to enable the security of future schemes to be more easily assessed.

\bibliographystyle{plain}
\bibliography{bib}

\appendix

\subsection{Proof of security for MFBE}
\label{proof}

\begin{proof} 
The proof goes by constructing a sequence of games:
\begin{description}[leftmargin=10pt]
    \item[Game 0] This game is defined as:  $$b\getsr\set{0,1};~ b'\get Exp^{\schemesymb,b}_{\adv}(1^\lambda);~ \mbox{return } b'$$ 
    Let $S_0$ be the event $b'=b$ in Game 0. Then we have:
    {\footnotesize
    \begin{align*}
Adv_\A^{\mfbe} & = \abs{\prob{\Exp_{\A}^{\mfbe, 1} = 1} - \prob{\Exp_{\A}^{\mfbe, 0} = 1}} \\
 & = \abs{\prob{\Exp_{\A}^{\mfbe, 1} = 1} + \prob{\Exp_{\A}^{\mfbe, 0} = 0} - 1} \\
 & = \abs{2\cdot\prob{S_0}-1}
\end{align*}}
    \item [Game 1] This game is the same as Game 0 except the decryption $DEC^{\sigma\texttt{-}ciph}$ in the oracle $\calO^{unwrap}$ is performed using a table lookup rather than real decryption operation. Formally, we modify the oracles  $\calO^{wrap}$ and  $\calO^{unwrap}$ as below: 
        \begin{align*}
        & \underline{\calO^{wrap}(pol):} \\
        &  k, w \get KGen(1^\lambda)\\
        &  (\phi, \overline{pol}) \get WrapKey^\sigma(k, w, pol) \\
        & L_{wrap} = L_{wrap}\cup \set{(k, w, \phi, \overline{pol})} \\
        & \mbox{return } (\phi, \overline{pol})
    \end{align*}
    The association between $k, w, \phi, \overline{pol}$ is recorded in a list $L_{wrap}$ which is initialised to be $\emptyset$. Correspondingly, we modify $\UnwrapKey^\sigma(\phi,\overline{pol})$ in the unwrap oracle $\calO^{unwrap}$ as below: 
    \begin{align*}
        & \underline{\UnwrapKey^\sigma(\phi,\overline{pol}):} \\
        & \mbox{If } (\cdot, w, \cdot, \overline{pol}) \in L_{wrap} \mbox{ for some } w \\
        & \quad ciph \get \overline{pol}.cipher \\
        & \quad  \mbox{{if} pol.usage = DeviceDataAfterBoot}\\ 
        & \qquad \mbox{or  pol.usage = UserDataAfterBoot} \mbox{ then} \\
        & \qquad k \get DEC^{ciph}(\phi, w)\\
        & \qquad \mbox{return } k \\
        &  \mbox{else return } \bot
    \end{align*}
    Let $S_1$ be the event $b'=b$ in Game 1.  Let $E$ be the event that the adversary queries $\calO^{unwrap(\cdot, \overline{pol})}$ such that $DEC^{ciph}(C, \sigma\texttt{-}key)\neq \bot$ and $C$ is not generated by any $\calO^{wrap}$ query, where $C  = \overline{pol}.wrapkey$. 
    Since Game 0 and Game 1 proceed identically unless $E$ occurs, we have $\abs{\prob{S_0}-\prob{S_1}}\leq \prob{E}$.
    
    Next we shall prove that $\prob{E} \leq Adv^{CTXT}_{\sigmaciph, \B_1}$.
    For any adversary $\A$ that attacks Game 1 and runs in time at most $\tau$, we can construct an adversary $\B_1^{{ENC_{\sigma-key}^{\sigmaciph}}, {DEC_{\sigma-key}^{\star, \sigmaciph}}}$ that attacks $\sigmaciph$ in the INT-CTXT sense and runs in time  at most $\tau$: $\B_1$ chooses $b\getsr \set{0,1}$ and runs $\A$ as a subroutine. $\B_1$ answers the oracle query $\calO^{wrap}$  queries using its oracle $ENC_{\sigma-key}^{\sigmaciph}$ and maintains a list of pairs of queried plaintext and the returned ciphertext. $\B_1$ answers the queries to $\calO^{unwrap}$ using the list to find the corresponding plaintext. For any ciphertext $C$ that is not in the list, $\B_1$ queries to  $DEC_{\sigma-key}^{\star, \sigmaciph}$. If $DEC_{\sigma-key}^{\star, \sigmaciph}(C) = 1$, $\B_1$ stops; otherwise $\B_1$ returns $\bot$. 
    Therefore we have $\abs{\prob{S_0}-\prob{S_1}}\leq Adv^{CTXT}_{\sigmaciph,\B_1}$.
    
    \vspace{10pt}
    \item[Game 2] This game is exactly the same as Game 1 except the wrap key encrypted with $\sigma\texttt{-}key$ is uniform random $w'$. Formally, the oracle $\calO^{wrap}$ is answered as below:
    \begin{align*}
        & \underline{\calO^{wrap}(pol):} \\
        & k, w, w' \get KGen(1^\lambda)\\
        & ciph \get pol.cipher \\
        & \overline{pol} \get pol \\
         &   \mbox{{if} pol.usage = DeviceDataAfterBoot}\\ 
        & \quad \mbox{or  pol.usage = UserDataAfterBoot} \mbox{ then} \\
         & \quad \phi \get ENC^{ciph}(k, w) \\
        & \quad \overline{pol}.wrapkey \get ENC^{\sigma\texttt{-}ciph}(w', \sigma\texttt{-}key) \\
        & \quad L_{wrap} = L_{wrap}\cup \set{(k, w, w', \phi, \overline{pol})} \\
        & \quad \mbox{return } (\phi, \overline{pol}) \\
         &  \mbox{else return } \bot
    \end{align*}
    The class key $k$ is encrypted with $w$, while the wrap key encrypted with $\sigma\texttt{-}key$ is actually $w'$.  The association between $k, w, w', \phi, \overline{pol}$ is recorded in a list $L_{wrap}$ which is initialised to be $\emptyset$. Correspondingly, we modify $\UnwrapKey^\sigma(\phi,\overline{pol})$ in the unwrap oracle $\calO^{unwrap}$ as below: 
        \begin{align*}
       & \underline{\UnwrapKey^\sigma(\phi,\overline{pol}):} \\
        & \mbox{If } (\cdot, w, \cdot, \cdot, \overline{pol}) \in L_{wrap} \mbox{ for some } w \\
        & \quad ciph \get \overline{pol}.cipher \\
        & \quad  \mbox{{if} pol.usage = DeviceDataAfterBoot}\\ 
        & \qquad \mbox{ or  pol.usage = UserDataAfterBoot} \mbox{ then} \\
        & \qquad k \get DEC^{ciph}(\phi, w)\\
        & \qquad \mbox{return } k \\
        &   \mbox{else return } \bot
    \end{align*}
    Let $S_2$ be the event $b'= b$. We shall prove that $\abs{\prob{S_2}-\prob{S_1}}\leq Adv_{\sigmaciph, \B_2}^{CPA}$. We construct the IND-CPA adversary $\B_2^{Enc^{\sigmaciph}_{\sigmakey}(LR(\cdot, \cdot, d))}$ with $d\in\set{0,1}$ that attacks $\sigmaciph$ as follows. $\B_2$ chooses $b\getsr\set{0,1}$ and runs Game 1. $\B_2$ answers the query to $\calO^{wrap}$ by choosing $k, w, w'\getsr KGen(1^\lambda)$, $\phi\get ENC^{ciph}(k,w)$ and queries its left-right oracle to get $\overline{pol}.wrapkey \get Enc^{\sigmaciph}_{\sigma\texttt{-}key}(LR(w, w', d))$. $\B_2$ records $(k,w,w',\phi,\overline{pol})$ in the list $L_{wrap}$. To answer queries to $\calO^{unwrap}$ on $(\phi, \overline{pol})$, $\B_2$ checks if $(\cdot, w, \cdot, \cdot, \overline{pol})\in L_{wrap}$. If so, $\B_2$ uses $w$ to decrypt $\phi$. Otherwise $\B_2$ returns $\bot$. All the other oracles can be easily answered since $\B_2$ has all the class keys $k$. $\B_2$ outputs $b'= b$.
    
    If $d=0$, $\B_2$ simulates Game 1  perfectly. If $d=1$, $\B_2$ simulates Game 2 perfectly. Therefore we have 
    {\footnotesize
       \begin{align*}
       Adv_{\sigmaciph,\B_2}^{CPA} & =  \abs{\prob{\Exp_{\sigmaciph, \B_2}^{CPA, 1} = 1} - \prob{\Exp_{\sigmaciph, \B_2}^{CPA, 0} = 1}} \\
       & = \abs{\prob{S_2} - \prob{S_1}}
          \end{align*}
  }
  
  \item[Game 3] This game is defined exactly the same as Game 2 except
    \begin{itemize}[leftmargin=10pt]
        \item Uniformly and randomly select $\eta \getsr [q_{w}]$ where $q_{w}$ is the total number of queries to $\calO^{wrap}$.
        \item Let $(k_\eta, w_\eta, w'_\eta, \phi_\eta, \overline{pol}_\eta)\in L_{wrap}$ be the values generated in the $\eta$-th query to $\calO^{wrap}$. In the challenge query, if the key corresponds to $kid^\star$ is not $k_\eta$, then aborts.
    \end{itemize}
    Let $S_3$ be the event $b'=b$. Clearly, the probability that abort does not happen is $1/q_{w}$ since $\eta$ is uniformly and randomly chosen. When abort does not happen, Game 3 is the same as Game 2. Therefore $\prob{S_3} = \prob{S_2}/q_{w}$.
    \item[Game 4] This game is defined exactly the same as Game 3 except:
    \begin{itemize}[leftmargin=10pt]
        \item Let $(k_\eta, w_\eta, w'_\eta, \phi_\eta, \overline{pol}_\eta)\in L_{wrap}$ be the values generated in the $\eta$-th query to $\calO^{wrap}$. When query $(\phi, \overline{pol}_\eta)$ to $\calO^{unwrap}$ for some $\phi$, replace the decryption $DEC^{ciph}(\phi, w_\eta)$ with the table lookup: if $\phi = \phi_\eta$ then get $k_{\eta}$ from $(k_\eta, \cdot, \cdot, \phi_\eta, \overline{pol}_\eta)\in L_{wrap}$; otherwise return $\bot$.
    \end{itemize}
    Let $S_4$ be the event $b'=b$ in Game 4. Let $E$ be the event that the adversary queries $\calO^{unwrap(\phi,\ov{pol}_{\eta})}$ such that $\phi\neq \phi_\eta$ and $DEC^{ciph}(\phi, w_\eta)\neq \bot$. Since Game 4 and Game 3 are exactly the same unless $E$ happens, we have $\abs{\prob{S_4} - \prob{S_3}}\leq \prob{E}$. We can prove that $\prob{E}\leq Adv_{ciph, \B_3}^{CTXT}$ for some adversary $\B_3$ that attacks $ciph$ in the INT-CTXT sense. The proof is similar to the analysis in Game 1 and is thus omitted. Based on this we can have  $\abs{\prob{S_4} - \prob{S_3}}\leq Adv_{ciph,\B_3}^{CTXT}$.
    
    \item[Game 5] This game is defined exactly the same as Game 4 except that the $\eta$-th query to $\calO^{wrap}$ is answered as below:
        \begin{align*}
        & \underline{\calO^{wrap}(pol_\eta):} \\
        & k_\eta, k'_\eta, w_\eta, w'_\eta \get KGen(1^\lambda)\\
        & ciph \get pol_\eta.cipher \\
        & \overline{pol}_\eta \get pol_\eta \\
        &   \mbox{{if} pol.usage = DeviceDataAfterBoot}\\ 
        & \quad \mbox{or  pol.usage = UserDataAfterBoot} \mbox{ then} \\
        & \quad \phi_\eta \get ENC^{ciph}(k'_\eta, w_\eta) \\
        & \quad  \overline{pol}_{\eta}.wrapkey \get ENC^{\sigmaciph}(w'_\eta, \sigmakey) \\
        & \quad L_{wrap} = L_{wrap}\cup \set{(k_\eta,k_\eta', w_\eta, w_\eta', \phi_\eta, \overline{pol}_\eta)} \\
         & \quad \mbox{return } (\phi_\eta, \overline{pol}_{\eta}) \\
          &  \mbox{else return } \bot
\end{align*}
    And we modify $\UnwrapKey^\sigma(\phi_\eta,\overline{pol}_\eta)$ in the unwrap oracle $\calO^{unwrap}$ on $(\phi_\eta, \overline{pol}_\eta)$  as below: 
        \begin{align*}
         & \underline{\UnwrapKey^\sigma(\phi_\eta,\overline{pol}_\eta):} \\
        & \mbox{If } (k_\eta, \cdot, \cdot, \cdot, \phi_\eta, \overline{pol}_\eta) \in L_{wrap} \\
        & \quad ciph \get \overline{pol}_\eta.cipher \\
        & \quad  \mbox{{if} pol.usage = DeviceDataAfterBoot}\\ 
        & \qquad \mbox{ or  pol.usage = UserDataAfterBoot} \mbox{ then} \\
        & \qquad\quad \mbox{return } k_\eta \\
        & \quad  \mbox{else return } \bot
    \end{align*}
    Note that the above modification is only for the $\eta$-th query, and the other queries are answered exactly the same as in Game 4. Let $S_5$ be the event $b'=b$. Similar to the analysis in Game 2, we can prove that        
    \begin{align*}
       Adv_{ciph,\B_4}^{CPA} & =  \abs{\prob{\Exp_{ciph, \B_4}^{CPA, 1}= 1} - \prob{\Exp_{ciph, \B_4}^{CPA, 0} = 1}} \\
       & = \abs{\prob{S_5} - \prob{S_4}}
          \end{align*}
          \item[Game 6] This game is exactly the same as Game 5 except the key derivation function $F_{k_\eta}: \set{0,1}^\ell \mapsto \set{0,1}^n$ is replaced with a random function $R$ obtained by $R \getsr Rand^{\ell\mapsto n}$, where $k_\eta$ is the class key generated in the $\eta$-th query to $\calO^{wrap}$ and $Rand^{\ell\mapsto n}$ is the family of all functions from $\set{0,1}^\ell$ to $\set{0,1}^n$. 
          Let $S_6$ be the event $b'=b$. We shall prove 
          $$\abs{\prob{S_6}-\prob{S_5}}\leq  Adv^{PRF}_{F,\B_5}$$
          Let $\calO_0 = R$ and $\calO_1 = F_{k_\eta}$. We construct the PRF adversary $\B_5^{\calO_d(\cdot)}$ with $d\in\set{0,1}$ as follows. $\B_5$ chooses $b\getsr\set{0,1}$ and runs Game 5. When $\B_5$ answers the $\eta$-th query to $\calO^{wrap}$, $\B_5$ sets $k_\eta = \bot$ which means $\B_5$ does not know the value of $k_\eta$.  $\B_5$ gets the values of $F_{k_\eta}(n_c)$ in the query $\calO^{enc}$ (resp. $\calO^{dec}$) on $(kid, \fnode)$ (resp. $(kid, \enode)$) with $kid$ being $k_\eta$'s handler by querying $\calO_d(n_c)$. When $k_\eta$ corresponds to the class key used in the challenge query which means abort does not happen, the adversary is not allowed to query $\calO^{reveal}(kid^\star)$. $\B_5$ outputs $b'=b$.
          
          If $d=1$, $\B_5$ simulates Game 5 perfectly. If $d=0$, $\B_5$ simulates Game 6 perfectly. Therefore we have 
              \begin{align*}
       Adv_{F,\B}^{PRF} & =  \abs{\prob{\Exp_{F, \B_5}^{PRF, 1} = 1} - \prob{\Exp_{F, \B_5}^{PRF, 0} = 1}} \\
       & = \abs{\prob{S_5} - \prob{S_6}}
          \end{align*}

Next we shall prove that 
\begin{align*}
    \abs{2\prob{S_6}-\frac{1}{q_w}} = Adv^{CCA}_{\fileciph, \B_6}
\end{align*} for some adversary $\B_6$ that attacks $\fileciph$ in the IND-CCA sense. Given an adversary $\A$ that attacks Game 6 and runs in time at most $\tau$, we  construct the adversary $\B_6^{\E_K(LR(\cdot,\cdot, b)), \D_K}$ that runs in time at most $\tau$ as follows. $\B_6$ chooses $\eta\getsr[q_{w}]$. For the $\eta$-th query to $\calO^{wrap}$, $\B_6$ sets $k_\eta = \bot$. The key derivation function $F_{k_\eta}$ is replaced with a random table $R$. For the challenge query $(kid^\star, \fnode_0, \fnode_1)$, if $kid^\star$ is not the key handler for $k_\eta$ then $\B_6$ aborts. Otherwise $\B_6$ chooses $n_c\getsr \set{0,1}^\ell$ and implicitly sets $R[n_c] = K$ since $\B_6$ does not know $K$. $\B_6$ creates the challenge ciphertext $\enode^\star$ by setting  $\enode^\star.meta = n_c$ and $\enode^\star.content \get \E_K(LR(\allowbreak \fnode_0, \fnode_1, b))$ obtained by querying $\B_6$'s left-right encryption oracle. To answer further encryption query to $\calO^{enc(kid^\star, \fnode)}$ with $\fnode.meta = n_c$, $\B_6$ queries 
$e\get \E_K(LR(\fnode,\allowbreak \fnode, b))$ and returns $\enode$ with $\enode.meta = n_c$ and $\enode.content = e$. The decryption query $\calO^{dec(kid^\star, \enode)}$ with $\enode.meta = n_c$ can be answered by querying $\D_K$ when $\enode$ is neither  $\enode^\star$ nor obtained from $\calO^{enc(kid^\star, \cdot)}$. The other queries can be easily answered and are omitted here because $\B$ selects all other class keys and file encryption keys. Finally $\A$ outputs $b'$ and $\B_6$ outputs $b'$. 

It is easy to see that $\B_6$ simulates Game 6 perfectly. Only when $\B_6$ does not abort, it can output 0/1. Since $\eta$ is uniformly and randomly chosen, the probability that $\B_6$ does not abort is $1/q_{w}$. Therefore,
{\footnotesize
              \begin{align*}
      & Adv_{\fileciph,\B_6}^{CCA}  \\
     & = \abs{\prob{\Exp_{\fileciph, \B_6}^{CCA, 1} = 1} - \prob{\Exp_{\fileciph, \B_6}^{CCA, 0} = 1}} \\
    & = \abs{\prob{\Exp_{\fileciph, \B_6}^{CCA, 1} = 1} - \frac{1}{q_w} + \prob{\Exp_{\fileciph, \B_6}^{CCA, 0} = 0}} \\
       & = \abs{2\prob{S_6} - \frac{1}{q_w}}
          \end{align*}
}

\end{description}
   Combing all the results above, we obtain    
   \begin{align*}
Adv&_\A^{\mfbe}  =  \abs{2\cdot\prob{S_0}-1} \\
& \leq 2 Adv_{\sigmaciph, \B_1}^{CTXT} + 2 Adv_{\sigmaciph, \B_2}^{CPA} + 2 q_w Adv_{ciph, \B_3}^{CTXT} \\
& \quad + 2 q_w Adv_{ciph, \B_4}^{CPA} +  2 q_w Adv_{F, \B_5}^{PRF} + q_w  Adv_{\fileciph, \B_6}^{CCA}\\
\end{align*}
This completes the proof.
\end{proof}

\subsection{Security definitions for symmetric ciphers}

\begin{definition}[IND-CPA and IND-CCA \cite{bellare1997concrete}]
Let $\calS\E = (\K, \E, \D)$ be a symmetric encryption scheme. Let $b\in\set{0,1}$ and $\A$ be an adversary. We define the left-right oracle $\calO^{\E_k(\calL\R(\cdot, \cdot, b))}$, to take input $(x_0, x_1)$ and return $\E_k(x_b)$, and we write $\calO^{\D_k(\cdot)}$ for the decryption oracle. Consider the following experiments:

{\footnotesize
\begin{tabular}{c|c}    
\multicolumn{2}{c}{}\\
    \begin{minipage}{0.42\columnwidth}
\[
\begin{array}{l}
 \Exp_{\calS\E, \A}^{CPA, b}(1^\lambda):  \\
   \quad k \getsr \K(1^\lambda)  \\
 \quad  b' \get \A^{\calO^{\E_k(\calL\R(\cdot, \cdot, b))}}(1^\lambda) \\
  \quad \mbox{return } b' 
\end{array}
\]
\end{minipage}
&
   \begin{minipage}{0.4\columnwidth}
\[
\begin{array}{l}
\Exp_{\calS\E, \A}^{CCA, b}(1^\lambda):  \\
\quad  k \getsr \K(1^\lambda) \\
 \quad  b' \get \A^{\calO^{\E_k(\calL\R(\cdot, \cdot, b))}, \calO^{\D_k(\cdot)}}(1^\lambda)\\
  \quad \mbox{return } b' 
\end{array}
\]
    \end{minipage}\\
    \multicolumn{2}{c}{}\\
\end{tabular}
}

In the CCA experiment, the adversary cannot query $\calO^{\D_k(\cdot)}$ on a ciphertext output by $\calO^{\E_k(\calL\R(\cdot, \cdot, b))}$. We define the advantages of the adversaries via 
{\footnotesize
\begin{align*}
Adv_{\calS\E, \A}^{CPA}(1^\lambda) & = \abs{\prob{\Exp_{\calS\E, \A}^{CPA, 1}(1^\lambda) = 1} - \prob{\Exp_{\calS\E, \A}^{CPA, 0}(1^\lambda) = 1}   } \\
Adv_{\calS\E, \A}^{CCA}(1^\lambda) & = \abs{\prob{\Exp_{\calS\E, \A}^{CCA, 1}(1^\lambda) = 1} - \prob{\Exp_{\calS\E, \A}^{CCA, 0}(1^\lambda) = 1}   }
\end{align*}
}
The scheme $\calS\E$ is said to be IND-CPA (resp. IND-CCA) secure if the advantage $Adv_{\calS\E, \A}^{CPA}(1^\lambda)$ (resp. $Adv_{\calS\E, \A}^{CCA}(1^\lambda)$) is negligible for any PPT adversary $\A$.
\end{definition}

\begin{definition}[Finite pseudorandom functions (PRFs)]
Given a family of functions $F: Keys(F)\times \set{0,1}^{\ell} \mapsto \set{0,1}^{n}$, we write $f\getsr F$ for the operations $K\getsr Keys(F), f\get F_K$, which means selecting a random function $f$ from the family $F$. We write $Rand^{\ell\mapsto n}$ the family of all functions from $\set{0,1}^\ell$ to  $\set{0,1}^n$. Let $b\in\set{0,1}$. Consider the following experiment:
\begin{align*}
 \Exp&_{F, \A}^{PRF, b}:  \\
   &  \calO_0 \getsr Rand^{\ell\mapsto n}; \calO_1 \getsr F \\
   &  b' \get \A^{\calO_b(\cdot)} \\
   & \mbox{return } b' 
\end{align*}
We define the advantage of the adversary via 
\begin{align*}
\footnotesize
Adv_{F, \A}^{PRF} & = \abs{\prob{\Exp_{F, \A}^{PRF, 1} = 1} - \prob{\Exp_{F, \A}^{PRF, 0} = 1}   }
\end{align*}
\end{definition}

\begin{definition}[Integrity of ciphertext (INT-CTXT) \cite{bellare2008authenticated}]
Let $\calS\E = (\K, \E, \D)$ be a symmetric encryption scheme. Let $b\in\set{0,1}$ and $\A$ be an adversary. Consider the following experiments:
\begin{align*}
\footnotesize
 \Exp&_{\calS\E, \A}^{CTXT}(1^\lambda):  \\
  &  k \getsr \K(1^\lambda); S \get \emptyset  \\
  &  \mbox{If } \A^{\calO^{\E_k(\cdot)}, \calO^{\D^\star_k(\cdot)}} \mbox{ queries to } \calO^{\D^\star_k(C)} s.t. \\
  & \quad \D_k(C) = M \neq \bot \\
  & \quad \mbox{and } \calO^{\E_k(\cdot)} \mbox{ never return } C \\
&   \mbox{then return 1, else return 0} 
\end{align*}
We define the advantage of the adversary via 
\begin{align*}
\footnotesize
Adv_{\calS\E, \A}^{CTXT}(1^\lambda) & = \prob{\Exp_{\calS\E, \A}^{CTXT}(1^\lambda) = 1} 
\end{align*}
The scheme $\calS\E$ is said to be INT-CTXT secure if the advantage $Adv_{\calS\E, \A}^{CTXT}(1^\lambda)$ is negligible for any PPT adversary $\A$.

\end{definition}

\subsection{Algorithms}
\label{app:algorithms}

\begin{algorithm}
    \footnotesize
	\SetKwProg{function}{function}{}{}
	\KwIn{$enode,kid$}
	\KwOut{$\textit{f}node$}
		
	\BlankLine
	\function{$\mathsf{\Decrypt}(enode,kid)$}{
		$k_m  \xleftarrow{} {} GetKeyFromKernelKeyring(kid)$\\
		\If {$k$}{
		    $n \xleftarrow{} \textit{e}node.meta$\\
		    
            \If {$enode.type = directory$}{
                $k_d \xleftarrow{} KDF(k_m, n, mode)$\\
                $\textit{f}node.children \xleftarrow{} DEC^\fileciph(enode.children, k_d)$\\
            }
            
            \BlankLine

		    \If {$enode.type = file$}{
                $k_d \xleftarrow{} KDF(k_m, n, mode)$\\
                $\textit{f}node.content \xleftarrow{} DEC^\fileciph(enode.content, k_d)$\\
            }
            \Return $\textit{f}node$
		} \Else {
            \Return $\bot$
		}
	}
	\caption{File decryption in EXT4 filesystems (Android \& *NIX systems)}
	\label{alg:decryptfbe}
\end{algorithm}

\begin{algorithm}
    \footnotesize
	\caption{$\ProvisionKey$ (Android \& iOS)}
	\label{alg:provisionkey}

	\SetKwProg{function}{function}{}{}
	\KwIn{$k$}
	\KwOut{$kid$}
	\BlankLine
    \function{$\ProvisionKey(k)$}{
   	    $kid \xleftarrow{} Load(k)$\\
   	 	\Return $kid$
    }
\end{algorithm}

\begin{algorithm}
    \footnotesize
	\caption{$\EvictKey$}
	\label{alg:evictkey}

	\SetKwProg{function}{function}{}{}
	\KwIn{$kid$}
	\KwOut{$\{1,\bot\}$}
	\BlankLine
    \function{$\EvictKey(kid)$}{
        $ret \xleftarrow{} Clear(kid)$\\
   	    \Return $ret$
    }
\end{algorithm}

\removelatexerror
\begin{algorithm}[H]
    \footnotesize
	\SetKwProg{function}{function}{}{}
	
	\KwIn{$econtent,k$}
	\KwOut{$content$}
	\BlankLine
	\function{$DEC^{\fileciph}(econtent, k)$}{
        \For {$eblock \in econtent$}{
        	$\mathsf{iv} \xleftarrow{} \mathsf{GetIV}(eblock,k.mode)$\\
            $cblock \xleftarrow{} CIPHER(eblock, k.\mathsf{val},\mathsf{iv})$\\
            $content \xleftarrow{} content \| cblock$ \\
        }
    }
	
	\caption{Node content decrypt (Android \& *NIX systems)}
	\label{alg:secdec}
\end{algorithm}

\vspace{250px}

\end{document}